\documentclass[aps,onecolumn,nopacs,nofootinbib,floatfix,superscriptaddress]{revtex4}

\usepackage{graphicx}
\usepackage{amsfonts}
\usepackage{amssymb}
\usepackage{amsbsy}
\usepackage{hyperref}
\usepackage{amsmath}
\usepackage{mathrsfs}
\usepackage{latexsym}
\usepackage{natbib}
\usepackage{bm}
\usepackage{subfigure} 
\usepackage{color}
\usepackage{wasysym}
\usepackage{mathbbol}
\usepackage{bigints}
\allowdisplaybreaks
\usepackage[normalem]{ulem}
\usepackage[dvipsnames]{xcolor}
\usepackage{multirow}
\usepackage{physics}
\usepackage{comment}
\usepackage{booktabs}
\usepackage{array}
\usepackage[T1]{fontenc}

\definecolor{napiergreen}{rgb}{0.16, 0.5, 0.0}

\renewcommand*{\l}{\lambda_{\star}}

\def\nn{\nonumber} 

\def\f{\frac}
\def\l{\left}
\def\r{\right}
\def\d{{\rm d}}

\def\exp{\mathrm{Exp}}
\def\erf{\mathrm{Erf}}
\def\erfc{\mathrm{Erfc}}

\begin{document}

\title{Radiative process of tripartite entangled probes in inertial motion}

\author{Subhajit Barman}
\email{subhajit.barman@physics.iitm.ac.in}
\affiliation{Centre for Strings, Gravitation and Cosmology,
Department of Physics, Indian Institute of Technology Madras, 
Chennai 600036, India}

\author{K. Hari}
\email{hari.k@iitb.ac.in}
\affiliation{Department of Physics, Indian Institute of Technology Bombay, Mumbai 400076, India}

\begin{abstract}

\noindent We study the radiative process of three entangled quantum probes initially prepared in a tripartite W state. As a basic set-up, we consider the probes to be inertial in flat spacetime and investigate how the radiative process is affected by different probe configurations. We take the quantum probes as either static or moving with uniform velocities and consider different switching scenarios. Our main observation confirms that the radiative process depends distinctively on the initial configuration in which the probes are arranged, as well as on the direction of the probe velocity. We also extend our analysis to a thermal environment, thereby simulating a more realistic background. We thoroughly discuss the effects due to different switchings, the thermal background, and probe motion on the radiative process of these tripartite entangled probes. We also comment on how the observations from this work can help prepare a set-up least affected by quantum decoherence.

\end{abstract}

\maketitle


\section{Introduction}\label{sec:introduction}

In the centenary of modern quantum mechanics, \textit{quantum entanglement} has emerged as one of its most profound and efficacious features. Quantum entanglement refers to the non-classical correlation between two or more particles that can be causally disconnected, a feature unique to a quantum system. Experimentally validated \cite{Freedman:1972zza, Kirby:2013hbs, Hensen:2015ccp}, entanglement has become imperative in the development of many technologies with enormous future prospects, such as quantum computing \cite{Feynman:1981tf, Aspect:1982fx, Jozsa:2002rcj}, quantum cryptography \cite{Ekert:1991zz, Jennewein:2000zz, Weihs:1998gy, Yin:2020rtd}, and quantum communication \cite{Tittel:1998ja, Horodecki:2009zz, Hu:2021uyl}. Furthermore, in recent times, specifically over the last two decades, a notable new development has emerged, popularly known as the field of Relativistic Quantum Information (RQI), which explores the interplay between quantum information and quantum field theory. 



RQI encompasses the necessary tools to analyze the phenomena involving quantum systems, non-inertial frames, and gravity, offering a crucial framework for understanding how relativistic effects and gravity can influence entanglement in quantum systems. There are several approaches and related set-ups in RQI that aim to quantify this influence. For instance, there are approaches that explore the possibility of transferring entanglement from the quantum field to suitable quantum probes \cite{Reznik:2002fz, Koga:2018the, Koga:2019fqh, Pozas-Kerstjens:2016rsh} with the intention of further utilizing that entanglement as a resource in quantum information experiments. The qualitative features in this induced entanglement vary significantly depending on the influence of gravity and the motion of the set-up \cite{Barman:2021bbw, Barman:2021kwg, Barman:2022xht, Barman:2023rhd, Barman:2023aqk, K:2023oon, Mayank:2025bkc}. At the same time, understanding the resilience of entanglement and mitigating its fragility - a phenomenon known as quantum decoherence - is vital to implement the entanglement as a practical resource for quantum information protocols \cite{Joos:1984uk, Zurek:1991vd, Schlosshauer:2014pgr}. Recent investigations have shed some light on decoherence due to gravity \cite{Das:2017qwp, Singh:2023ppv, Moustos:2024ymw, Takeda:2025cye} and relativistic effects arising from the relative motion of entangled systems \cite{Shresta:2003, Bernad:2013, Pikovski:2015wwa}. There are also extensive studies investigating the radiative process of bipartite entangled systems \cite{Arias:2015moa, Menezes:2015iva, Picanco:2020api, Barman:2021oum, Barman:2022utm, Barman:2024vah}, which, resembling certain aspects of decoherence, are equally sensitive to the motion of the set-ups and the background. This emphasizes the necessity of a deeper understanding of how relative motion and gravity can influence quantum correlations.\vspace{0.1cm}

While a significant part of the literature has focused on bipartite entangled systems due to their relative simplicity, the future of quantum technology, relevant to RQI, lies in multipartite entanglement \cite{Membrere:2023vao, Wu:2024jgy, Huang:2025yvj, Rufo:2025egf, Li:2025bzd, Hamzehofi:2025, Park:2025}. A key reason is that the only configuration that can be produced from two entangled probes is a straight line, whereas with at least three probes, triangular configurations of different characteristics can be considered. At the same time, tripartite entanglement offers more possible initial states compared to bipartite states, thus presenting a broader scope for implementing quantum information protocols and preparing experimental set-ups. In particular, the seminal work of Greenberger, Horne, and Zeilinger (GHZ) \cite{greenberger1990bell}, which provided a compelling experimental test of quantum non-locality, utilized a tripartite entangled state (now known as the GHZ state). With the emergence of technologies in quantum communications and distributed quantum computing, the need to study multipartite entanglement has increased even more. We believe they will offer new avenues for quantum protocols, presenting unique challenges for maintaining coherence and providing insights into overcoming those challenges.\vspace{0.1cm}

In this article, in a first of its kind, we investigate the radiative process of a tripartite entangled system initially prepared in a W state. Our primary focus is to analyze how this process is affected by the spatial orientation and relative motion of an entangled system of three parties. We model each party as a two-level atom (qubit) interacting with a background quantum field. The fluctuations of this background field induce the spontaneous de-excitation of the atoms to an uncorrelated ground state, leading to the degradation of entanglement. Moreover, we extend our analysis to examine the effects as the surrounding environment transitions to a thermal state, thereby simulating a more realistic and noisy environment. We study the collective transition probability rates for transitions from different initial entangled states, which provide insights into the dynamics of tripartite entanglement in relativistic settings and facilitate understanding the qualitative features of entanglement degradation.\vspace{0.1cm}

This manuscript is organized in the following manner. In Sec. \ref{sec:formulation}, we introduce different tripartite entangled states and identify the specific W states \cite{Dur:2000zz} relevant for our purpose. We also sketch the model set-up in terms of two-level atoms as a particle probe, and elucidate the measure to quantify the degradation from the entangled state. In Sec. \ref{sec:radiative-static}, we start our analysis with different configurations of static probes and investigate the effects of spatial orientation in the decay of entanglement with and without a switching function that controls the interaction of the atoms with the quantum field. Subsequently, in Sec. \ref{sec:radiative-vel}, the effect of relative motion in the radiative process is considered. To model a noisy environment, in Sec. \ref{sec:thermal-bath}, we consider the radiative process with a quantum field in a thermal bath. Finally, we conclude with our observations and insights in the Sec. \ref{sec:discussion}.\vspace{0.1cm}

\textbf{Notation}: We work in natural units, $c=\hbar=1$ with metric signature $(-,+,+,+)$. The boldface for distance and velocity ${\bf d},\,{\bf v}$ is used to denote 3-vector quantities. The symbol $\lfloor ~ \rfloor$ is used to indicate the floor function, and `mod 2' is an abbreviation for modulo 2. The double arrow underset is used to indicate a symmetric sum, i.e,  $R_{\underleftrightarrow{jl}} = R_{jl} + R_{lj}$. 

\section{Formulation: model set-up}\label{sec:formulation}
In this section, we provide a brief introduction to different tripartite entangled states. We also present the model set-up necessary for investigating the radiative process of quantum probes prepared in suitable tripartite entangled states.

\subsection{Tripartite entangled states}\label{subsec:tripartite-states}

Most discussions on entanglement are around simple, but non-trivial cases of bipartite entanglement. Even though the bipartite systems are a convenient model for understanding the foundations of Quantum Information (QI), multipartite entanglement provides new opportunities to develop new protocols in QI. They are relevant for quantum communication or computing between multiple users. The first step towards multipartite entanglement is tripartite entanglement. In the case of bipartite entanglement, usually, the entangled states (pure state) are represented in Bell basis.
\begin{subequations}
\begin{align}
    \ket{{\varphi}^{\pm}} &= \frac{1}{\sqrt{2}}\left(\ket{g_Ag_B}\pm\ket{e_Ae_B}\right), \\ 
    \ket{\psi^{\pm}} &= \frac{1}{\sqrt{2}}\left(\ket{g_Ae_B}\pm\ket{e_Ag_B}\right)~.
    \label{eq:BP-states}
\end{align}
\end{subequations}
where $\ket{g}$ and $\ket{e}$ represents the ground state and excited state respectively. The above states can be transformed to any other basis by applying local unitary transformations and classical communication (LOCC), hence only one class of entanglement. Unlike the bipartite entangled states, the tripartite (or any multipartite) entanglement there can be different classes of entanglement. The two inequivalent classes of entanglement between parties for tripartite entanglement \cite{Cunha:2019jex} (parties $A$, $B$ and $C$) are signified by the Greenberger-Horne-Zeilinger (GHZ) states \cite{greenberger1990bell, Greenberger:1989tfe} and W states \cite{Dur:2000zz}, given by,
\begin{subequations}\label{eq:TP-states}
\begin{eqnarray}
    \ket{\text{GHZ}}&=&\frac{1}{\sqrt{2}}\left( \ket{g_Ag_Bg_C} + \ket{e_Ae_Be_C} \right) \\
    \ket{\text{W}}&=&\frac{1}{\sqrt{3}}\left( \ket{g_Ag_Be_C}+\ket{g_Ae_Bg_C}+\ket{e_Ag_Bg_C} \right)~.\label{eqn:W-state-1}
\end{eqnarray}
\end{subequations}
When the GHZ states are expressed with an arbitrary amount of entanglement, the states are defined using,
\begin{eqnarray}\label{eq:GHZ-states-gen}
    \ket{\text{GHZ}_{g_Ag_Bg_C}}=\cos\vartheta\ket{g_Ag_Bg_C} + \sin\vartheta\ket{e_Ae_Be_C}~,
\end{eqnarray}
and with this, the GHZ basis can be constructed from local operations (LOCC) on $\ket{\text{GHZ}_{ggg}}$.
Similarly, the W basis can be constructed from local operations (LOCC) on $\ket{\text{W}_{ggg}}$. The family of W states \cite{Schwemmer:2015qbi} can be expressed as,
\begin{widetext}
\begin{subequations}\label{eq:W-states-all}
\begin{eqnarray}
&&\ket{\text{W}_{1}}=\sin\vartheta\,\cos\chi\ket{gge}+\sin\vartheta\,\sin\chi \ket{geg} +\cos\vartheta \ket{egg}~, \\
&&\ket{\text{W}_{2}}=\sin\vartheta\,\sin\chi\ket{gge}-\sin\vartheta\,\cos\chi \ket{geg} +\cos\vartheta \ket{eee}~, \\
&&\ket{\text{W}_{3}}=-\sin\vartheta\,\sin\chi\ket{egg}+\sin\vartheta\,\cos\chi \ket{eee} +\cos\vartheta \ket{geg}~, \\
&&\ket{\text{W}_{4}}=\sin\vartheta\,\cos\chi\ket{egg}+\sin\vartheta\,\sin\chi \ket{eee} +\cos\vartheta \ket{gge}~, \\
&&\ket{\text{W}_{5}}=\sin\vartheta\,\cos\chi\ket{eeg}+\sin\vartheta\,\sin\chi \ket{ege} +\cos\vartheta \ket{gee}~, \\
&&\ket{\text{W}_{6}}=\sin\vartheta\,\sin\chi\ket{eeg}-\sin\vartheta\,\cos\chi \ket{ege} +\cos\vartheta \ket{ggg}~, \\
&&\ket{\text{W}_{7}}=-\sin\vartheta\,\sin\chi\ket{gee}+\sin\vartheta\,\cos\chi \ket{ggg} +\cos\vartheta \ket{ege}~, \\
&&\ket{\text{W}_{8}}=\sin\vartheta\,\cos\chi\ket{gee}+\sin\vartheta\,\sin\chi \ket{ggg} +\cos\vartheta \ket{eeg}~;
\end{eqnarray}
\end{subequations}
\end{widetext}
where, $0\le\vartheta\le\pi/2$, $0\le\chi\le\pi/2$, and the subscripts $A,\,B\,\text{and}\,C$ are avoided to reduce the clutter, but the leftmost position in subscript in the state definition like $\ket{\text{X}_{ggg}}$ is for particle $A$, the central position is for $B$ and the rightmost position denote particle $C$, where `X' is any of the state in Eq. \ref{eqn:W-state-1}. In the next section, the three qubits modelled as particle probes will be considered in the W state. The GHZ basis will not be considered since the states in this basis are not eigenstates of the probe Hamiltonian to be considered in the following analysis.

\subsection{Model set-up}\label{subsec:model-setup}

In this section, we provide a model for investigating the radiative process of three Unruh-DeWitt probes that are initially in specific tripartite entangled state. These are atomic two-level probes that interact with the background quantum field and were initially conceptualized to understand the Unruh and the Hawking effects \cite{Unruh:1976db, hawking1975}. Later on, they became an important tool to realize different quantum entanglement phenomena mediated through the background field \cite{Reznik:2002fz, Salton:2014jaa, Pozas-Kerstjens:2015gta, Arias:2015moa, Koga:2018the, Picanco:2020api, Zhou:2020oqa, Barman:2021oum, Barman:2021bbw, Barman:2021kwg, Barman:2022utm, Barman:2022xht, Barman:2023rhd, Barman:2023aqk, K:2023oon, Barman:2024vah, Mayank:2025bkc}. In particular, we consider three Unruh-deWitt probes to be denoted by $A$, $B$, and $C$ respectively. We also consider that these probes interact with the background massless, minimally coupled quantum scalar field $\Phi(x)$ through linear monopole interactions, see \cite{Barman:2021oum} for a similar set-up. The entire Hamiltonian for this probe field system is given by 
\begin{eqnarray}\label{eq:total-Hamiltonian-1}
    \mathcal{H} = \mathcal{H}_{D}+\mathcal{H}_{F}+\mathcal{H}_{I}~.
\end{eqnarray}
Here $\mathcal{H}_{D}$ denotes the free probe Hamiltonian, $\mathcal{H}_{F}$ denotes the field Hamiltonian, and $\mathcal{H}_{I}$ represents the interaction between the probe and the field. The expressions for these different components of the Hamiltonian are given by 
\begin{subequations}\label{eq:total-Hamiltonian-2}
\begin{eqnarray}
    \mathcal{H}_{D} &=& \frac{\omega_{0}}{2}\sum_{j=A}^{C}\Big(|e_{j}\rangle\langle e_{j}|-|g_{j}\rangle\langle g_{j}|\Big)~,\\
    \mathcal{H}_{F} &=& \frac{1}{2}\int\,d^3x\,\big[\dot{\Phi}^2(x)+|\nabla\Phi(x)|^2\big]~,\\
    \mathcal{H}_{I} &=& \sum_{j=A}^{C} \mu_{j}\,m_{j}(\tau_{j})\,\kappa_{j}(\tau_{j})\,\Phi[x_{j}(\tau_{j})]~.
\end{eqnarray}
\end{subequations}
In the above expression, $|g_{j}\rangle$ and $|e_{j}\rangle$ denote the ground and excited states of the $j^{th}$ probe. $\omega_{0}$ denotes the energy gap between the ground and excited states of a certain probe. It is to be noted that we have considered identical probes and thus this energy gap is the same for all probes. The overhead dot in $\dot{\Phi}(x)$ denotes the time derivative. Moreover, $\mu_{j}$, $m_{j}$, and $\kappa_{j}$ respectively denote the interaction strength, the monopole moment operator, and the switching function. Here, one can easily show that none of the GHZ states are eigenstates of the detector Hamiltonian $\mathcal{H}_{D}$.

We consider that initially, the probe system is in a collective state $|\bar{\omega}\rangle$ and the field is in the Minkowski ground state $|0_{M}\rangle$. After time evolution the entire system will be in a state $\mathcal{U}~|\bar{\omega},\,0_{M}\rangle$, where $\mathcal{U}$ denotes the time evolution operator. For identical interaction between the probes and the field $\mu_{1}=\mu_{2}=\mu_{3}$, say the interaction strength is now $\mu$, this time evolution operator can be obtained as
\begin{eqnarray}\label{eq:time-translation-opp}
    \mathcal{U} &=& \mathcal{T}~\Bigg\{\exp\bigg[-i\,\mu\int_{-\infty}^{\infty}\,\sum_{j=A}^{C} m_{j}(\tau_{j})\,\kappa_{j}(\tau_{j})\,\Phi[x_{j}(\tau_{j})]\,d\tau_{j}\bigg]\Bigg\}~. \nn \\
\end{eqnarray}
To consider a transition of the probe system to reach a state $\ket{\widetilde{\omega}}$ after the time evolution, we need to find the transition probability. Let us consider that the probe and field is initially in $\ket{\bar{\omega}}$ and vacuum $\ket{0_{M}}$ respectively, and later probe will be in state $|\widetilde{\omega}\rangle$ and the field will be in some arbitrary state $|\Psi\rangle$. Then the transition amplitude for this transition from the state $|\bar{\omega},\,0_{M}\rangle$ to $|\widetilde{\omega},\,\Psi\rangle$ will be
\begin{widetext}
\begin{eqnarray}\label{eq:transition-amp}
    \mathcal{A}_{|\bar{\omega},\,0_{M}\rangle\to |\widetilde{\omega},\,\Psi\rangle} &=& \langle\Psi,\,\widetilde{\omega}|~\hat{\mathcal{U}}~|\bar{\omega},\,0_{M}\rangle~,\nonumber\\
    &\approx& -i\,\mu\,\langle\Psi,\,\widetilde{\omega}|~\int_{-\infty}^{\infty}\,\sum_{j=A}^{C} m_{j}(\tau_{j})\,\kappa_{j}(\tau_{j})\,\Phi[x_{j}(\tau_{j})]\,d\tau_{j}~|\bar{\omega},\,0_{M}\rangle~,
\end{eqnarray}
\end{widetext}
considering the interaction strength $\mu$ to be perturbatively small and taking terms upto $\mathcal{O}(\mu)$. The transition probability for the transition from the probe state $|\bar{\omega}\rangle$ to $|\widetilde{\omega}\rangle$ will be
\begin{eqnarray}\label{eq:transition-prob}
    \Gamma(\omega)=\Gamma_{|\bar{\omega}\rangle\to |\widetilde{\omega}\rangle} &=& \sum_{\{|\Psi\rangle\}}\mathcal{A}_{|\bar{\omega},\,0_{M}\rangle\to |\widetilde{\omega},\,\Psi\rangle}\mathcal{A}_{|\bar{\omega},\,0_{M}\rangle\to |\widetilde{\omega},\,\Psi\rangle}^{*}~,\nonumber\\
    &=& \mu^2\,\sum_{j,l=A}^{C} m_{j}^{\widetilde{\omega}\bar{\omega}^*} m_{l}^{\widetilde{\omega}\bar{\omega}^*}\,F_{jl}(\omega)~,
\end{eqnarray}
where, $\omega=\widetilde{\omega}-\bar{\omega}$, denotes the energy difference between the two states $|\widetilde{\omega}\rangle$ and $|\bar{\omega}\rangle$. Moreover, $m_{j}^{\widetilde{\omega}\bar{\omega}}=\langle \widetilde{\omega}|~m_{j}(0)~|\bar{\omega}\rangle$ denotes the expectation value of the monopole moment operator in the probe initial and final energy states. In particular, the expression of the monopole moment operator is $m_{j}(0)=|e_{j}\rangle\langle g_{j}|+|g_{j}\rangle\langle e_{j}|$. The quantities $F_{jl}(\omega)$ are similar in form to the response functions. Their form is given by 
\begin{widetext}
\begin{eqnarray}\label{eq:response-fn}
    F_{jl}(\omega) = \int_{-\infty}^{\infty}d\tau'_{l}\int_{-\infty}^{\infty}d\tau_{j}\,\kappa_{j}(\tau_{j})\,\kappa_{l}(\tau'_{l})\,e^{-i\,\omega(\tau_{j}-\tau'_{l})}\,G_{jl}^{+}\big[x_{j}(\tau_{j}),\, x_{l}(\tau'_{l})\big]~,
\end{eqnarray}
\end{widetext}
with $G_{jl}^{+}[x_{j}(\tau_{j}),\, x_{l}(\tau'_{l})]$ denoting the positive frequency Wightman function between the field modes at $x_{j}(\tau_{j})$ and $x_{l}(\tau'_{l})$. From the expression of Eq. \eqref{eq:response-fn}, one can observe that $F_{jj}$ denotes the single probe response functions, while $F_{jl}$ with $j\neq l$ is due to the correlation between different probes. We respectively call them auto and cross-transition probabilities. Here, we would also like to mention that for certain switching functions, it is convenient to deal with the rates of these transition probabilities. In particular, we define the generalized rate of these transition probabilities as
\begin{subequations}\label{eq:Rjl-Def}
\begin{eqnarray}\label{eq:Rjl-Def-a}
    R_{jl}(\omega) &=& \frac{F_{jl}(\omega)}{\tilde{T}}~\\
    \textup{and,~~} \mathcal{R}(\omega) &=& \frac{1}{\mu^2}\frac{\Gamma(\omega)}{\tilde{T}}~;\label{eq:Rjl-Def-b}
\end{eqnarray}
\end{subequations}
It should be noted that for eternal switching $\kappa(\tau)=1$, the above denominator $\tilde{T}=\lim_{T\to\infty}(1/2)\int_{-T}^{T}ds$. At the same time, for Gaussian switching the above quantity is $\tilde{T}=(1/2) \int_{-\infty}^{\infty}ds\, e^{-s^2/(2T^2)} = T\,\sqrt{\pi/2}$, see \cite{Barman:2022utm}. We should also mention that the rate defined in Eq. \eqref{eq:Rjl-Def-a} corresponds to the individual auto and cross transition probabilities. At the same time, the rate defined in Eq. \eqref{eq:Rjl-Def-b} correspond to the collective transition probability.

\vspace{0.2cm}

\begin{figure*}
\includegraphics[width=12cm]{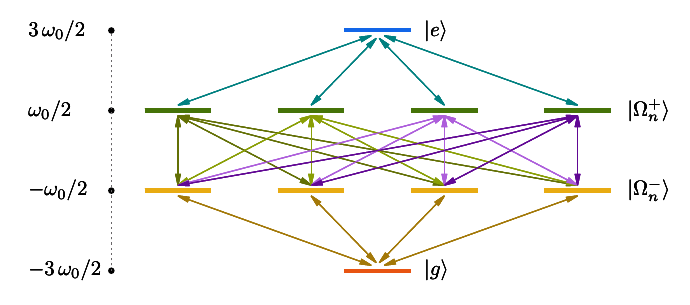}
\caption{In the above figure, we have depicted the energy levels of the collective quantum states for a system of three entangled atomic probes through a schematic diagram. All the permissible transitions between different energy levels are also highlighted in this figure. As previously mentioned, one can note that the separable states, i.e., the collective ground $|g\rangle$ and excited $|e\rangle$ states, have no degeneracies. At the same time, the entangled states $|\Omega^{-}_{n}\rangle$ are degenerate in their energy eigenvalues. Similarly, the entangled states $|\Omega^{+}_{n}\rangle$ are also degenerate in their energy eigenvalues. From this figure, one can notice that the transitions from the state $|e\rangle$ to $|\Omega^{-}_{n}\rangle$ and $|g\rangle$ are not possible and vice versa. Similarly, the transition from the state $|\Omega^{+}_{n}\rangle$ to $|g\rangle$ and its reverse are not possible.}
    \label{fig:energy-levels}
\end{figure*}

In the scenario, when we have three probes, there can be four collective eigen-states of the probe Hamiltonian $\mathcal{H}_{D}$. These states are respectively $|g\rangle$, $|\Omega^{-}_{n}\rangle$, $|\Omega^{+}_{n}\rangle$, and $|e\rangle$ with energies $(-3\omega_{0}/2)$, $(-\omega_{0}/2)$, $\omega_{0}/2$, and $3\omega_{0}/2$. In particular, the states $|g\rangle=|g_{A}\rangle \otimes|g_{B} \rangle\otimes|g_{C}\rangle$ and $|e\rangle=|e_{A}\rangle\otimes|e_{B} \rangle\otimes|e_{C}\rangle$ are non-degenerate and denote the collective ground and the excited states of the three probe system. One can also notice that these states, $|g\rangle$ and $|e\rangle$, are separable in terms of individual probe states. At the same time, the states $|\Omega^{-}_{n}\rangle$ and $|\Omega^{+}_{n}\rangle$ contain four degenerate eigen-states each. We included the $n$ in their subscript to denote four different values corresponding to these four degenerate states. 
The states $|\Omega^{-}_{n}\rangle$ and $|\Omega^{+}_{n}\rangle$ are given by the compact form,
\begin{widetext}
\begin{subequations}
\begin{align}
    |\Omega^{-}_{n}\rangle &=\frac{1}{\sqrt{3}}\Big[|g_{A}\rangle\otimes|g_{B}\rangle\otimes|e_{C}\rangle + (-1)^{b_1}\,|g_{A}\rangle\otimes|e_{B}\rangle\otimes|g_{C}\rangle + (-1)^{b_2}\,|e_{A}\rangle\otimes|g_{B}\rangle\otimes|g_{C}\rangle\Big]~, \\
    |\Omega^{+}_{n}\rangle &=\frac{1}{\sqrt{3}}\Big[|e_{A}\rangle\otimes|e_{B}\rangle\otimes|g_{C}\rangle + (-1)^{b_1}\,|e_{A}\rangle\otimes|g_{B}\rangle\otimes|e_{C}\rangle + (-1)^{b_2}\,|g_{A}\rangle\otimes|e_{B}\rangle\otimes|e_{C}\rangle\Big]~,
\end{align}
\end{subequations}
\end{widetext}
where
\begin{subequations}\label{eq:binary-decomposition}
\begin{align}
    b_1 &= (n-1) \bmod 2~,  \\
    b_2 &= \left\lfloor \frac{n-1}{2} \right\rfloor~;
\end{align}
\end{subequations}
with $\lfloor ~ \rfloor$ signifying the floor function.

One can obtain the expectation values of the monopole moment operators for the transition from $|\Omega^{-}_{1}\rangle$ to $|g\rangle$ as $m_{j}=(1/\sqrt{3}),\,(1/\sqrt{3}),\,(1/\sqrt{3})$, where the three values respectively correspond to $j$ being $A$, $B$, or $C$ signifying different probes. For the transition from $|\Omega^{-}_{2}\rangle$ to $|g\rangle$ as $m_{j}=(1/\sqrt{3}),\,(-1/\sqrt{3}),\,(1/\sqrt{3})$, and for the transition from $|\Omega^{-}_{3}\rangle$ to $|g\rangle$ as $m_{j}=(1/\sqrt{3}),\,(1/\sqrt{3}),\,(-1/\sqrt{3})$. Finally for the transition from $|\Omega^{-}_{4}\rangle$ to $|g\rangle$ as $m_{j}=(1/\sqrt{3}),\,(-1/\sqrt{3}),\,(-1/\sqrt{3})$. The energy levels and the permissible transitions among them for three entangled probes are depicted in Fig. \ref{fig:energy-levels}.
With the help of these monopole moment expectation values and Eq. \eqref{eq:Rjl-Def-b} one can get the general expressions for the collective transition probability rates for transitions from $|\Omega^{-}_{n}\rangle$ to $|g\rangle$ as
\begin{align}\label{eq:Rtotal-OmTg-gen}
    \mathcal{R}_{|\Omega^{-}_{n}\rangle\to |g\rangle}(\omega) = \frac{1}{3} \Big[ & \sum_{j=A}^{C} R_{jj} + (-1)^{b_1} R_{\underleftrightarrow{AB}} + (-1)^{b_2} R_{\underleftrightarrow{AC}} \nn\\
    & + (-1)^{b_1+b_2} R_{\underleftrightarrow{BC}} \Big]~,
\end{align}
where $R_{\underleftrightarrow{jl}} = R_{jl} + R_{lj}$ and $b_1,b_2$ are same as defined before.

In our subsequent analysis, we shall use these generalized expressions to obtain the final forms of the collective transitions corresponding to different scenarios.

\section{Radiative process of static quantum probes}\label{sec:radiative-static}

In this section, we consider probe configurations where all the probes are static and investigate the radiative process. In particular, we consider a generalized static three-probe set-up, i.e., the positions of the probes can be from coaxial to the vertices of a triangle with variable arm length. We would like to mention that we will consider both the eternal and Gaussian switching scenarios for the understanding of the radiative process. In the subsequent analysis, we will first evaluate the auto and cross-transition probabilities, i.e., $F_{jj}$ and $F_{jl}$ respectively. Then we will talk about the collective transition probabilities. 

\begin{figure*}
\includegraphics[width=10cm]{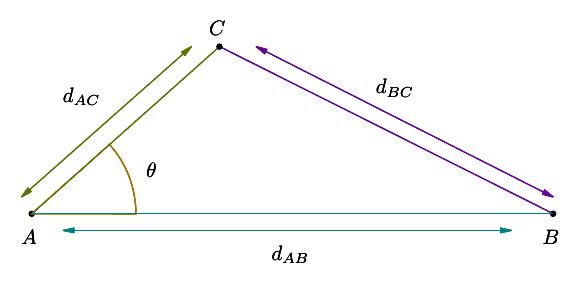}
\caption{We present a schematic diagram for three probes $A$, $B$, and $C$ that are static. The distances between these probes are respectively $d_{AB}$, $d_{BC}$, and $d_{AC}$. The angle formed between $d_{AB}$ and $d_{AC}$ is $\theta$. By varying these parameters of three distances and the angle one gets all possible probe configurations.}
    \label{fig:static-detectors}
\end{figure*}

First, we consider three static probes as depicted in Fig. \ref{fig:static-detectors}. The distances between the probes are parametrized by $d_{AB}$, $d_{AC}$, and $d_{_{BC}}$. In a similar fashion one can also parametrize these distances in terms of $d_{AB}$, $d_{AC}$, and the angle $\theta$, where the previous $d_{_{BC}}$ is given by $d_{BC}=(d_{AB}^2+d_{AC}^2-2d_{AB}d_{AC}\cos{\theta})^{-1/2}$. We also consider the probes to be in the entangled state $|\Omega^{-}_{n}\rangle$, and find the transition probabilities for the transition to the collective ground state $|g\rangle$. The probe trajectories are given by
\begin{subequations}\label{eq:SP-trajectories}
\begin{eqnarray}
    t_{A} &=& \tau_{A}\,,~{\bf x}_{A} = 0\,;\\
    t_{B} &=& \tau_{B}\,,~{\bf x}_{B} = {\bf d}_{AB}\, = d_{AB} \,{\hat{{\bf d}}_{B}};\\
    t_{C} &=& \tau_{C}\,,~{\bf x}_{C} = {\bf d}_{AC}\, = d_{AC} \, {\hat{{\bf d}}_{C}}.
\end{eqnarray}
\end{subequations}
where $\hat{\bf d}$ is the corresponding unit vector. 
%
%
For a single static probe, the Wightman function in Eq. \eqref{eq:SP-trajectories} for $j=l$ is given by
\begin{eqnarray}\label{eq:SP-Gjj}
    G_{jj}^{+}(\tau_{j},\tau'_{j}) &=& -\frac{1}{4\pi^2}\,\frac{1}{(\tau_{j}-\tau'_{j}-i\epsilon)^2} \nn \\
    &=& -\frac{1}{4\pi^2}\,\frac{1}{(u_{j}-i\epsilon)^2}~,
\end{eqnarray}
where, $u_{j}=\tau_{j}-\tau'_{j}$ whereas, when $j\neq l$, the Wightman function for static probes will be 
\begin{eqnarray}\label{eq:SP-Gjl}
    G_{jl}^{+}(\tau_{j},\tau'_{l}) &=& -\frac{1}{4\pi^2}\,\frac{1}{(\tau_{j}-\tau'_{l}-i\epsilon)^2-d_{jl}^2} \nn \\ 
    &=& -\frac{1}{4\pi^2}\,\frac{1}{(u_{jl}-i\epsilon)^2-d_{jl}^2}~,
\end{eqnarray}
where, $u_{jl}=\tau_{j}-\tau'_{l}$, and $d_{jl}$ is the separation between the $j^{th}$ and $l^{th}$ probe.

From the above expressions, we observed that both the Wightman functions are time translational invariant. Thus, they were written in a form reflective of this invariance, considering a change of variables to either $u_{j}$ or $u_{jl}$. In our subsequent analysis, we will be using these expressions to obtain the auto and cross transition probabilities.

\subsection{Eternal switching}\label{subsec:radiative-static-eternal}

In this part, we consider the eternal switching scenario with the window function given by $\kappa(\tau_{j})=1$, i.e., the probes interact with the background field for infinite time. Then with the help of Eqs. \eqref{eq:response-fn} and \eqref{eq:Rjl-Def-a}, and for $j=l$ the rate of auto-correlation $R_{jj}(\omega)$, which denote the rates of individual probe transition probabilities, will be 
\begin{eqnarray}\label{eq:SP-Rjj-eternal}
    R_{jj}(\omega) &=& -\frac{1}{4\pi^2}\int_{-\infty}^{\infty}du_{j}\frac{e^{-i\omega u_{j}}}{(u_{j}-i\epsilon)^2}~\nonumber\\
    ~&=& -\Theta(-\omega)\,\frac{\omega}{2\pi}~.
    \label{eqn:ind-rate-eternal}
\end{eqnarray}
On the other hand, when $j\neq l$, again we take the help of Eqs. \eqref{eq:response-fn} and \eqref{eq:Rjl-Def-a}. Then the rate of $F_{jl}(\omega)$ that denotes the cross correlation term will be 
\begin{eqnarray}\label{eq:SP-Rjl-eternal}
    R_{jl}(\omega) &=& -\frac{1}{4\pi^2}\int_{-\infty}^{\infty}du\frac{e^{-i\omega u}}{(u-i\epsilon)^2-d_{jl}^2}~\nonumber\\
    ~&=& -\frac{\Theta(-\omega)}{2\pi\,d_{jl}}~\sin{(\omega\,d_{jl})}~.
\end{eqnarray}
We should note that in the limit of $d_{jl}\to 0$ one shall get back the $R_{jj}(\omega)$ result from Eq. \eqref{eq:SP-Rjj-eternal} using the limit $\lim_{x\to 0}\,\textup{sinc}\,{x}=1$ of the sinc $(=\sin{x}/x)$ function.

\begin{figure*}
\centering
    \includegraphics[width=7.8cm]{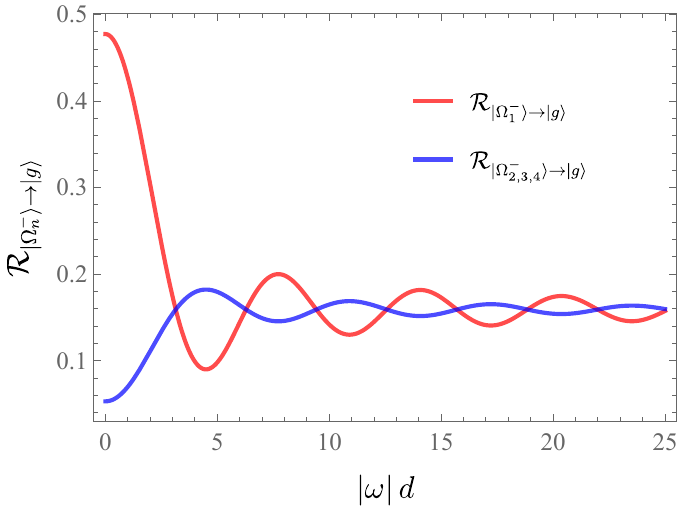}
    \hskip 30pt
    \includegraphics[width=7.8cm]{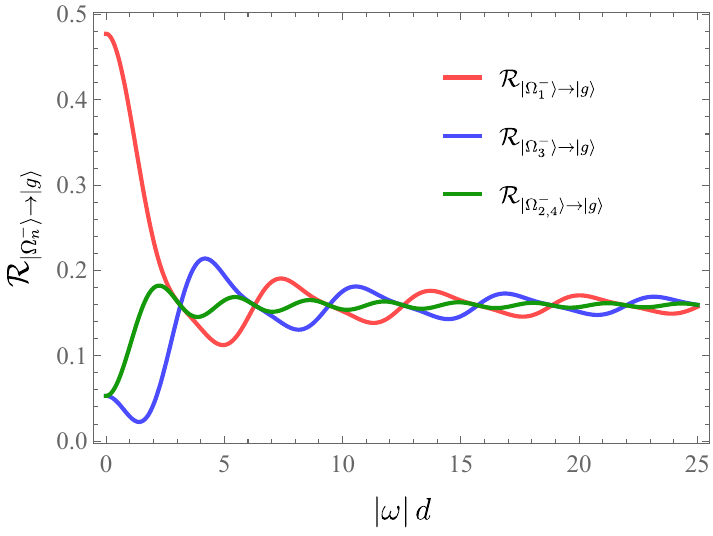}
    \vskip 20pt
    \includegraphics[width=7.8cm]{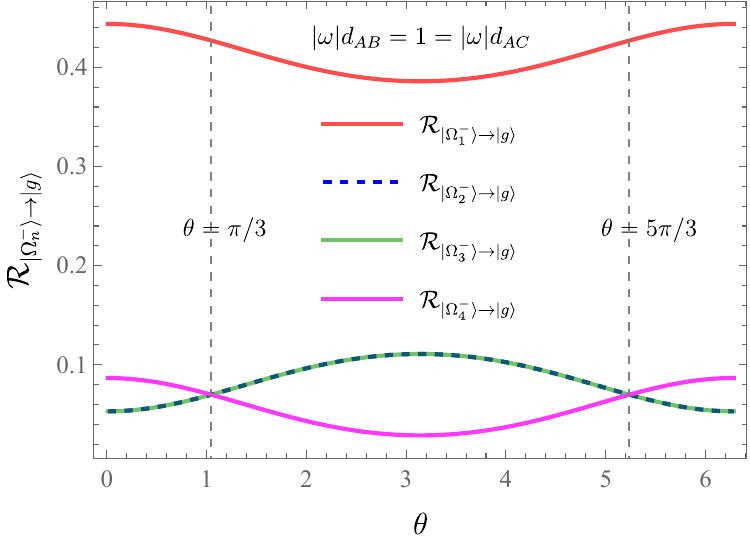}
    \hskip 30pt
    \includegraphics[width=7.8cm]{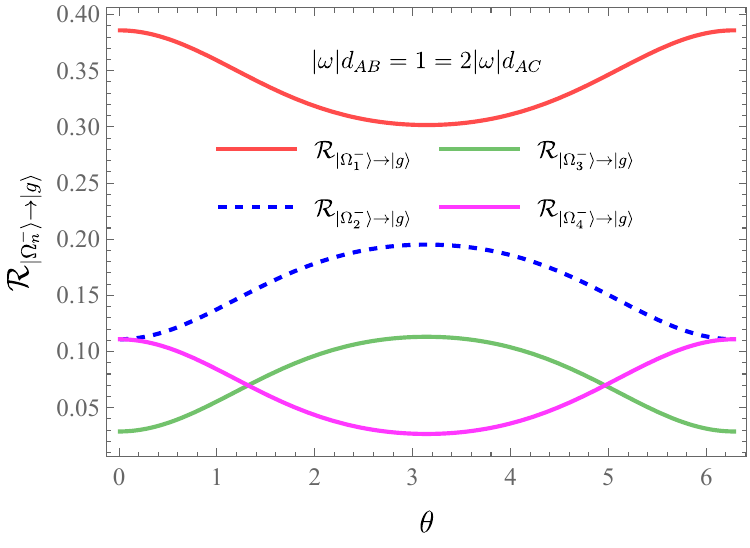}
    \caption{
    \textbf{Top-left:} We have plotted the collective transition probability rates $\mathcal{R}_{|\Omega^{-}_{n}\rangle\to|g\rangle}(\omega)$ as functions of the dimensionless probe energy gap $|\omega|d$ for a scenario where the three probes are stationed at the three vertices of an equilateral triangle. In this scenario, the transition rates from the states $|\Omega^{-}_{2}\rangle$, $|\Omega^{-}_{3}\rangle$, and $|\Omega^{-}_{4}\rangle$ are the same. \textbf{Top-right:} We have plotted the collective transition probability rates $\mathcal{R}_{|\Omega^{-}_{n}\rangle\to|g\rangle}(\omega)$ as functions of the dimensionless probe energy gap $|\omega|d$ when the probes are stationed equidistant on a straight line. In this scenario, we have considered $d_{AB}=2\,d_{AC}$, and here the transition probability rates from the states $|\Omega^{-}_{2}\rangle$ and $|\Omega^{-}_{4}\rangle$ are the same. From these plots, one can observe that as the energy gap or the separation between the probes increases, the different transition probability rates tend to reach a fixed value. \textbf{Bottom-left:} The collective transition probabilities are plotted as functions of the angle $\theta$ between the vectors ${\bf d}_{AB}$ and ${\bf d}_{AC}$ for $|\omega|\,d_{AB}=1=|\omega|\,d_{AC}$. From this plot, one can confirm that when $\theta=\pi/3$ or $\theta=5\pi/3$, i.e., when the probes are at the vertices of an equilateral triangle, the transition probability rates from the states $|\Omega^{-}_{2}\rangle$, $|\Omega^{-}_{3}\rangle$, and $|\Omega^{-}_{4}\rangle$ become equal. \textbf{Bottom-right:} The collective transition probabilities are plotted as functions of the angle $\theta$ between the vectors ${\bf d}_{AB}$ and ${\bf d}_{AC}$ for $|\omega|\,d_{AB}=1=2\,|\omega|\,d_{AC}$. From this plot, we confirm that when $\theta=0$ or $\theta=2\pi$, i.e., when the probes are on a straight line placed equidistant, the transition probability rates from the states $|\Omega^{-}_{2}\rangle$ and $|\Omega^{-}_{4}\rangle$ become equal. We would like to mention that all of the above plots in this figure are obtained for eternal switching $\kappa(\tau_{j})=1$.
    }
    \label{fig:R-st-et-vOm}
\end{figure*}

\subsubsection{Collective transition probability rates and their characteristics}\label{subsubsec:radiative-st-et-discussion}

In this part of the section, we present the expressions for the collective transition probability rates for transitions from the state $|\Omega^{-}_{n}\rangle$ to $|g\rangle$, i.e., for de-excitations to the ground state from the excited state $|\Omega^{-}_{n}\rangle$. It is to be noted that there are no transitions permissible from the collective excited state $|e\rangle$ to the collective ground state $|g\rangle$, or from the states $|\Omega^{+}_{n}\rangle$ to $|g\rangle$. With the help of Eqs. \eqref{eq:Rjl-Def-a} and \eqref{eq:Rtotal-OmTg-gen}, one can get the rate of these collective transition probabilities. In particular, the total transition rates for transitions from different $|\Omega^{-}_{n}\rangle$ states to the ground state $|g\rangle$ will be in the compact form,
\begin{widetext}
\begin{align}\label{eq:Rtotal-st-et}
    \mathcal{R}_{\ket{\Omega^{-}_{n}} \to \ket{g}} =-\frac{\Theta(-\omega)}{2\pi}\,\Big[\omega + \frac{2}{3} \left\{(-1)^{b_1} \, \frac{\sin{(\omega\,d_{AB})}}{d_{AB}} + (-1)^{b_2} \, \frac{\sin{(\omega\,d_{AC})}}{d_{AC}} + (-1)^{b_1 + b_2} \, \frac{\sin{(\omega\,d_{BC})}}{d_{BC}} \right\}\Big]
\end{align}
\end{widetext}
where, $b_1$ and $b_2$ is the same definition as given in Eq. \eqref{eq:binary-decomposition}.

In the above expressions of the transition probabilities, let us consider $\theta$ to be the angle between the vectors ${\bf d}_{AB}$ and ${\bf d}_{AC}$, representing the displacement vectors of the probes $B$ and $C$ from $A$. One can also visualize this configuration from Fig. \ref{fig:static-detectors}. In this scenario, one can express the distance $d_{BC}$ as $d_{BC}=(d_{AB}^2+d_{AC}^2-2d_{AB}d_{AC}\cos{\theta})^{1/2}$. Then, it is to be noted that one can move from one probe configuration to another by changing $\theta$.

For probes sitting on the vertices of an equilateral triangle, we have $d_{AB}=d_{AC}=d_{BC}$. We denote this separation to be $d$, i.e., $d_{AB}=d_{AC}=d_{BC}=d$. In this scenario, the above expressions get simplified. In particular, we will have the collective transition probability rates to be
\begin{subequations}
\begin{align}
    \mathcal{R}_{|\Omega^{-}_{1}\rangle\to |g\rangle}(\omega) &= -\frac{\Theta(-\omega)}{2\pi}\,\Big[\omega+\frac{2\,\sin{(\omega\,d)}}{d}\Big]~,\\
    \mathcal{R}_{|\Omega^{-}_{2,3,4}\rangle\to |g\rangle}(\omega) &= 
     -\frac{\Theta(-\omega)}{2\pi}\,\Big[\omega-\frac{2}{3}\,\frac{\sin{(\omega\,d)}}{d}\Big]~.
     \label{eq:Rtotal-st-et-EqLat}
\end{align}
\end{subequations}
where, $\mathcal{R}_{|\Omega^{-}_{2,3,4}\rangle\to |g\rangle}(\omega) \implies \mathcal{R}_{|\Omega^{-}_{2}\rangle\to |g\rangle}(\omega) = \mathcal{R}_{|\Omega^{-}_{3}\rangle\to |g\rangle}(\omega) = \mathcal{R}_{|\Omega^{-}_{4}\rangle\to |g\rangle}(\omega)$.
One can observe that in this scenario, the transition from the states $|\Omega^{-}_{2}\rangle$, $|\Omega^{-}_{3}\rangle$, and $|\Omega^{-}_{4}\rangle$ are equal. In Fig. \ref{fig:R-st-et-vOm}, top-left, we have plotted the collective transition probability rates when the probes are sitting on the vertices of an equilateral triangle. From this figure, we observe that as the magnitude of the probe energy gap or the separation between the probes increases, all the transition probabilities tend to reach a single fixed value.

In Fig. \ref{fig:R-st-et-vOm}, top-right, we have also plotted the collective transition probability rates when the probes are kept equidistant on a straight line. In this scenario, we have considered $2d_{AC}=2d_{BC}=d_{AB}=2d$ and $\theta=0$, i.e., the probe $C$ is placed middle of the probes $A$ and $B$ on a straight line. We obtain the expressions for the transition probability rates as
\begin{widetext}
\begin{subequations}\label{eq:Rtotal-st-et-StLine}
\begin{eqnarray}
    \mathcal{R}_{|\Omega^{-}_{1}\rangle\to |g\rangle}(\omega) &=& -\frac{\Theta(-\omega)}{2\pi}\,\Big[\omega+\frac{4}{3}\,\frac{\sin{(\omega\,d)}}{d}+\frac{1}{3}\,\frac{\sin{(2\omega\,d)}}{d}\Big]~,\\
    \mathcal{R}_{|\Omega^{-}_{2,4}\rangle\to |g\rangle}(\omega) &=& -\frac{\Theta(-\omega)}{2\pi}\,\Big[\omega-\frac{1}{3}\,\frac{\sin{(2\omega\,d)}}{d}\Big] ~, \\
    ~\mathcal{R}_{|\Omega^{-}_{3}\rangle\to |g\rangle}(\omega) &=& -\frac{\Theta(-\omega)}{2\pi}\,\Big[\omega-\frac{4}{3}\,\frac{\sin{(\omega\,d)}}{d}+\frac{1}{3}\,\frac{\sin{(2\omega\,d)}}{d}\Big]~.
\end{eqnarray}
\end{subequations}
\end{widetext}
Here, the transitions from the states $|\Omega^{-}_{2}\rangle$ and $|\Omega^{-}_{4}\rangle$ are equal, which is also reflected in the plots from Fig. \ref{fig:R-st-et-vOm}. From this figure one can notice another feature that the transitions from the states $|\Omega^{-}_{2}\rangle$, $|\Omega^{-}_{3}\rangle$, and $|\Omega^{-}_{4}\rangle$ become equal when $d$ goes to zero, which can be corroborated by taking the limit $d\to 0$ in the expressions of Eq. \eqref{eq:Rtotal-st-et-StLine}. In Fig. \ref{fig:R-st-et-vOm}, bottom-left and bottom-right, we have plotted the collective transition probability rates as functions of the angle $\theta$. On left we have considered $d_{AB}=d_{AC}$ to reproduce the observations of equilateral triangle configuration when $\theta=\pi/3$ or $\theta=5\pi/3$. On the right, we have considered $d_{AB}=2d_{AC}$ to reproduce the observations of equidistant on a straight line configuration when $\theta=0$. Our observations suggest that as one moves from one probe configuration to another, the qualitative features in the collective transition probability rates corresponding to different entangled states change, which is evident from our plots with respect to $\theta$. An important point to note is that, even though the $\ket{\Omega_{n}^{-}}$ state is degenerate, this degeneracy is not reflected in the transition probabilities from $\ket{\Omega_{n}^{-}}$ to the ground state and it depends on the configuration of the probes. In other words, the probe configuration determines whether the transition probability from each transition can be distinguished in the radiative process.

\subsection{Finite Gaussian switching}\label{subsec:radiative-static-gaussian}

In this part of the section, we investigate the radiative process of static probes that interact with the background field through Gaussian switching function, i.e., $\kappa(\tau)=e^{-\tau^2/T^2}$. In this regard, we shall utilize the Wightman functions as mentioned in Eqs. \eqref{eq:SP-Gjj} and \eqref{eq:SP-Gjl}. The evaluation of the auto-transition probabilities $F_{jj}(\omega)$, and more specifically the rate of them $R_{jj}(\omega)$, is straightforward, and is available in the literature, see \cite{LSriramkumar_1996}. Therefore, we will not provide the derivation in the main text of the work, and we have provided it to the Appendix, see Appendix \ref{Appn:Fjj-Inertial-Gaussian}. In particular, for Gaussian switching the auto transition rate can be obtained to be
\begin{eqnarray}\label{eq:SP-Rjj-Gaussian}
    R_{jj}(\omega) &=& \frac{1}{(2\pi)^{3/2}T}\,\Big[e^{-\omega^2\,T^2/2}-\sqrt{\frac{\pi}{2}}\,T\,\omega\, \erfc\left( \frac{\omega \,T}{\sqrt{2}}\right)\Big]~, \nn \\
\end{eqnarray}
where we have utilized the definition of transition rate from Eq. \ref{eq:Rjl-Def-a} and the expression of the auto-transition probability from Eq. \eqref{eq:SP-Fjj-Gaussian-2}.

On the other hand, in a similar manner, one can proceed to find the cross transition probabilities. In particular, with the help of Eq. \eqref{eq:response-fn} one can get the $F_{jl}(\omega)$ for Gaussian switching as
\begin{eqnarray}\label{eq:SP-Fjl-Gaussian}
    F_{jl}(\omega) = -\frac{1}{4\pi^2}\int_{-\infty}^{\infty}d\tau_{j}\int_{-\infty}^{\infty}d\tau_{l}\,\frac{e^{-i\omega(\tau_{j}-\tau_{l})} e^{-\frac{(\tau_{j}^2+\tau_{l}^2)}{T^2}}}{(\tau_{j}-\tau_{l}-i\epsilon)^2-d_{jl}^2}~, \nn \\
\end{eqnarray}
We consider a change of variables $u_{jl}=\tau_{j}-\tau_{l}$, and $s_{jl}=\tau_{j}+\tau_{l}$, as well as the definition of the transition probability rate from Eq. \eqref{eq:Rjl-Def-b}. Then after integrating over $s_{jl}$ in Eq. \eqref{eq:SP-Fjl-Gaussian}, we shall get the expression of $R_{jl}(\omega)$ as
\begin{eqnarray}\label{eq:SP-Fjl-Gaussian-2}
    R_{jl}(\omega) = -\frac{1}{4\pi^2\,T\sqrt{2\pi}} \int_{-\infty}^{\infty}du_{jl}\,\frac{e^{-i\,\omega\, u_{jl}} e^{-u_{jl}^2/(2T^2)}}{(u_{jl}-i\epsilon)^2-d_{jl}^2}~. 
\end{eqnarray}
In the previous expression, we have utilized the integral representation of Eq. \eqref{eq:Gen-IntRep-1a}. Whereas, the $u_{jl}$ integration can be conducted by representing the Gaussian function as $e^{-u_{jl}^2/(2T^2)} = (T/\sqrt{2\pi}) \int_{-\infty}^{\infty}d\xi\,e^{i\,\xi\,u_{jl}-\xi^2\,T^2/2}$. Then, the final result looks like 
\begin{eqnarray}\label{eq:SP-Fjl-Gaussian-3}
    R_{jl}(\omega) = -\frac{i}{8\pi\, d_{jl}}\,e^{-i\omega\, d_{jl}-d_{jl}^2/(2T^2)}\bigg[\erfc\left(\frac{\omega T}{\sqrt{2}}-\frac{i\,d_{jl}}{\sqrt{2}T}\right) \nn \\ -e^{2i\,\omega d_{jl}}\,\erfc\left(\frac{\omega T}{\sqrt{2}}+\frac{i\,d_{jl}}{\sqrt{2}T}\right)\bigg]~.
\end{eqnarray}
This expression of $R_{jl}(\omega)$ in the limit of $d_{jl}\to 0$ results in the exact expression of $R_{jj}(\omega)$ from Eq. \eqref{eq:SP-Rjj-Gaussian} of the previous analysis. It is to be noted that the rate $R_{jl}(\omega)=F_{jl}(\omega)/(T\sqrt{\pi/2})$ is obtained from $F_{jl}(\omega)$ by dividing it by the $s_{jl}$ integral part, which is $(1/2) \int_{-\infty}^{\infty}ds_{jl}\, e^{-s_{jl}^2/(2T^2)}=(T\sqrt{\pi/2})$.

\begin{figure*}
\includegraphics[width=7.8cm]{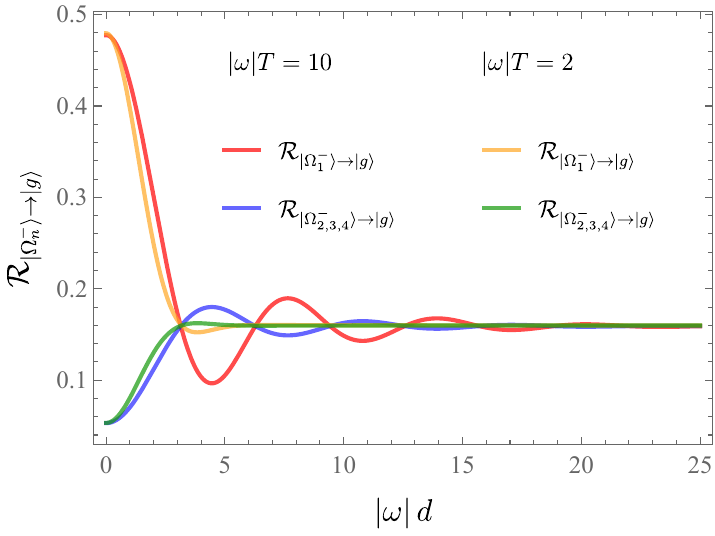}
\hskip 30pt
\includegraphics[width=7.8cm]{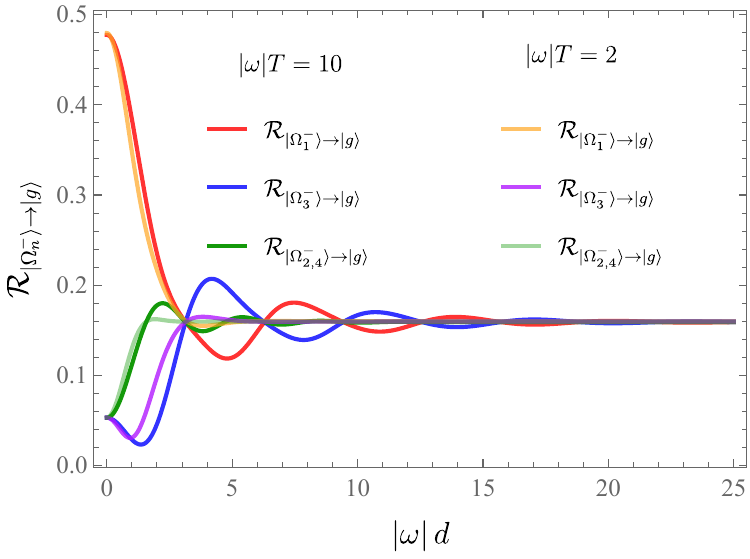}
\vskip 10 pt
\includegraphics[width=7.8cm]{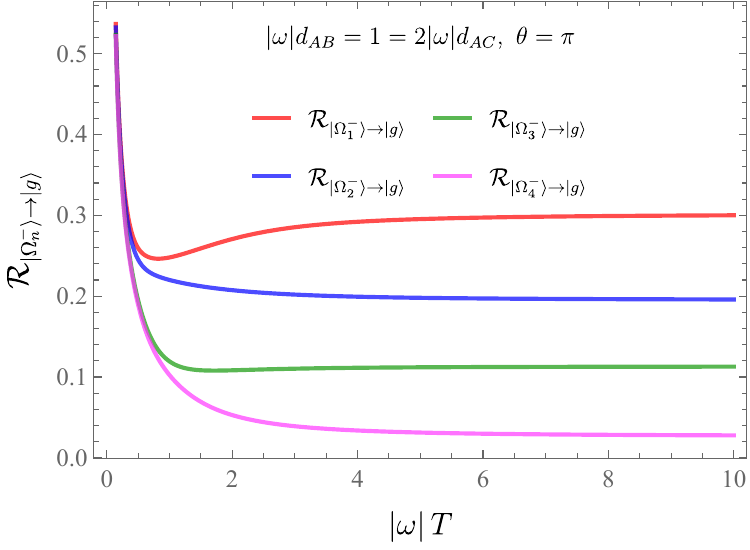}
\hskip 30pt
\includegraphics[width=7.8cm]{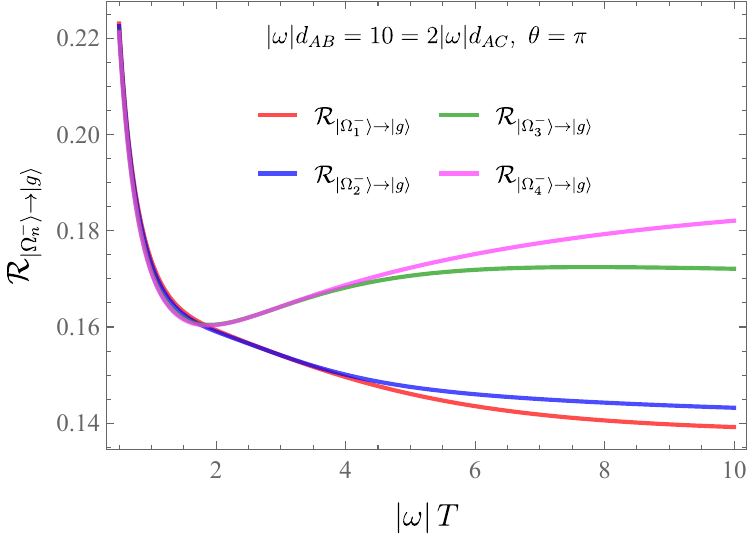}
\caption{\textbf{Top-left:} We have plotted the collective transition probability rates $\mathcal{R}_{|\Omega^{-}_{n}\rangle\to|g\rangle}(\omega)$ as functions of the dimensionless probe energy gap $|\omega|d$ for a scenario where the three probes are stationed at the three vertices of an equilateral triangle and for Gaussian switching. Here also, like the eternal switching scenario, the transition rates from the states $|\Omega^{-}_{2}\rangle$, $|\Omega^{-}_{3}\rangle$, and $|\Omega^{-}_{4}\rangle$ are equal. We have provided two sets of plots corresponding to fixed $|\omega|T=10$ and $|\omega|T=2$. One can observe that with decreasing $T$, the damping-like effect in the collective transition rate is larger. \textbf{Top-right:} We have plotted the collective transition probability rates as functions of the dimensionless probe energy gap $|\omega|d$ when the probes are stationed equidistant on a straight line. We have considered $d_{AB}=2\,d_{AC}$, and here the transition probability rates from the states $|\Omega^{-}_{2}\rangle$ and $|\Omega^{-}_{4}\rangle$ are the same. From these plots, one can also observe that as $T$ increases, we encounter fewer ripples in the collective transition probability rate, and all these rates reach a single fixed value for a smaller value of $d$ as compared to the situation of larger $T$. \textbf{Bottom-left:} The collective transition probability rates are plotted as functions of the dimensionless switching time $|\omega|T$ for $|\omega|d_{AB}=1=2|\omega|d_{AC}$ and $\theta=\pi$. \textbf{Bottom-right:}  The collective transition probability rates are plotted as functions of the dimensionless switching time $|\omega|T$ for $|\omega|d_{AB} =10= 2|\omega|d_{AC}$ and $\theta=\pi$. From the bottom-left and bottom-right plots, we observe that for very low switching time $(|\omega|T<1)$, all the transition probability rates increase with decreasing $T$. We also observe that for $|\omega|T\ge 2$, this feature is not the same for all the probability rates. All of these above plots in this figure are obtained for Gaussian switching of $\kappa(\tau_{j})=e^{-\tau_{j}^2/T^2}$.}
    \label{fig:R-st-Gsn-vOm}
\end{figure*}

\subsubsection{Collective transition probability rates and their characteristics}\label{subsubsec:radiative-st-gsn-discussion}

In this part of the section, we provide the expressions for the collective transition probability rates and investigate their characteristics for the Gaussian switching scenario. By putting the expressions of $R_{jj}(\omega)$ and $R_{jl}(\omega)$ from Eqs. \eqref{eq:SP-Rjj-Gaussian} and \eqref{eq:SP-Fjl-Gaussian-3} in Eq. \eqref{eq:Rtotal-OmTg-gen} one can get the expressions for the collective transition probability rates for transitions from $|\Omega^{-}_{n}\rangle$ to $|g\rangle$. Like the eternal switching scenario, here also one can observe that for probes fixed on the vertices of an equilateral triangle of arm length $d$, the transitions from the states $|\Omega^{-}_{2}\rangle$, $|\Omega^{-}_{3}\rangle$, and $|\Omega^{-}_{4}\rangle$ are equal. In particular, for Gaussian switching their expressions will be
\begin{widetext}
\begin{subequations}\label{eq:Rtotal-st-gsn-EqLat}
\begin{eqnarray}
    \mathcal{R}_{|\Omega^{-}_{1}\rangle\to |g\rangle}(\omega) &=& \frac{1}{(2\pi)\,T}\,\bigg[\frac{1}{\sqrt{2\pi}}\Big\{e^{-\omega^2\,T^2/2}-\sqrt{\frac{\pi}{2}}\,T\,\omega\, \erfc\left( \frac{\omega \,T}{\sqrt{2}}\right)\Big\}\nonumber\\
    ~&& -\frac{i}{2\, d}\,e^{-i\omega\, d-d^2/(2T^2)}\Big\{\erfc\left(\frac{\omega T}{\sqrt{2}}-\frac{i\,d}{\sqrt{2}T}\right)-e^{2i\,\omega d}\,\erfc\left(\frac{\omega T}{\sqrt{2}}+\frac{i\,d}{\sqrt{2}T}\right)\Big\}\bigg]~,\\
    \mathcal{R}_{|\Omega^{-}_{2}\rangle\to |g\rangle}(\omega) &=& \frac{1}{(2\pi)\,T}\,\bigg[\frac{1}{\sqrt{2\pi}}\Big\{e^{-\omega^2\,T^2/2}-\sqrt{\frac{\pi}{2}}\,T\,\omega\, \erfc\left( \frac{\omega \,T}{\sqrt{2}}\right)\Big\}\nonumber\\
    ~&& +\frac{i}{6\, d}\,e^{-i\omega\, d-d^2/(2T^2)}\Big\{\erfc\left(\frac{\omega T}{\sqrt{2}}-\frac{i\,d}{\sqrt{2}T}\right)-e^{2i\,\omega d}\,\erfc\left(\frac{\omega T}{\sqrt{2}}+\frac{i\,d}{\sqrt{2}T}\right)\Big\}\bigg]~\nonumber\\
    ~&=& 
    \mathcal{R}_{|\Omega^{-}_{3}\rangle\to |g\rangle}(\omega) =
    \mathcal{R}_{|\Omega^{-}_{4}\rangle\to |g\rangle}(\omega) ~.
\end{eqnarray}
\end{subequations}
\end{widetext}
In Fig. \ref{fig:R-st-Gsn-vOm}, top-left, we have plotted this collective transition probability rates as functions of the dimensionless energy gap $|\omega|d$, for probes on the vertices of an equilateral triangle. Our observations suggest that with Gaussian switching the ripples in these probability rates become lesser and lesser as one decreases the switching time $T$. This observation is also verified in the top-right plot of Fig. \ref{fig:R-st-Gsn-vOm}, where we have plotted the transition probability rates for probes equidistant on a straight line. It is to be noted that in the latter scenario, the transitions from the states $|\Omega^{-}_{2}\rangle$ and $|\Omega^{-}_{4}\rangle$ are equal, much like the eternal switching scenario. In Fig. \ref{fig:R-st-Gsn-vOm}, bottom-left and bottom-right, we have also plotted the transition probability rates as functions of the dimensionless switching time $|\omega|\,T$. From these plots we observed that for very low switching time $(|\omega|T<1)$, all the transition probability rates increase with decreasing $T$. We also observe that for $|\omega|T\ge 2$, this feature is not the same for all the probability rates.

\section{Radiative process of probes in inertial motion}\label{sec:radiative-vel}

In a situation for probes moving with uniform velocities, their relative motions will play a crucial role in providing different results compared to the previous static case. In this regard, we consider probe $A$ to be static and the other two probes in relatively uniform velocities. If $\tau_{j}$ be the proper time corresponding to the $j^{th}$ probe then these probe trajectories are described by
\begin{subequations}\label{eq:UVP-trajectories}
\begin{eqnarray}
    t_{A} &=& \tau_{A}\,,~{\bf x}_{A} = 0\,;\\
    t_{B} &=& \frac{\tau_{B}}{\sqrt{1-\textsc{v}_{B}^2}}\,,~{\bf x}_{B} = {\bf d}_{AB}+\frac{\textbf{\textsc{v}}_{B}\,\tau_{B}}{\sqrt{1-\textsc{v}_{B}^2}}\,;\\
    t_{C} &=& \frac{\tau_{C}}{\sqrt{1-v_{C}^2}}\,,~{\bf x}_{C} = {\bf d}_{AC}+\frac{\textbf{\textsc{v}}_{C}\,\tau_{C}}{\sqrt{1-v_{C}^2}}\,.
\end{eqnarray}
\end{subequations}
Here we have considered that probe $B$ and $C$ respectively have velocities $\textbf{\textsc{v}}_{B}$ and $\textbf{\textsc{v}}_{C}$. At $\tau_{B}=0$, probe $B$ is stationed at distance $d_{AB}$ on the $x-$axis. On the other hand, at $\tau_{C}=0$, the probe $C$ is stationed at a distance ${\bf d}_{AC}$ from $A$, which is a constant vector.

The different Wightman functions relevant for the estimation of the radiative process can be obtained from the generic form,
\begin{eqnarray}\label{eq:WightmanFn-gen}
    G^{+}_{jl} &=& -\frac{1}{4\pi^2} \frac{1}{(t_{j}-t_{l}-i\,\epsilon)^2-|{\bf x}_{j}-{\bf x}_{l}|^2}~.
\end{eqnarray}
When $j=l$, the spatial separation between the probes will become $|{\bf x}_{j}-{\bf x}'_{j}|=\textsc{v}_{j}\,\gamma_{j}(\tau_{j}-\tau'_{j})$. Naturally, for probe $A$ this quantity vanishes as the velocity is zero in this case. We have the spatial separation between the probes $A$ and $B$ as $|{\bf x}_{B}-{\bf x}_{A}|^2 =x_{B}^2+x_{A}^2-2\,{\bf x}_{B}.\,{\bf x}_{A}= d_{AB}^2 + \textsc{v}_{B}^2 \, \gamma_{B}^2\, \tau_{B}^2+2\,\gamma_{B}\,\tau_{B}\,{\bf d}_{1}.\textbf{\textsc{v}}_{B}$. The spatial separation between probe $A$ and $C$ is obtained in a similar manner and is the same as the previous expression with only the difference of subscript $B$ replaced by $C$. On the other hand, the spatial separation between probes $B$ and $C$ is $|{\bf x}_{B}-{\bf x}_{C}|^2 =|{\bf d}_{AB}-{\bf d}_{AC}|^2+|\gamma_{B}\,\tau_{B}\,\textbf{\textsc{v}}_{B}-\gamma_{C}\,\tau_{C}\,\textbf{\textsc{v}}_{C}|^2+2\,({\bf d}_{AB}-{\bf d}_{AC}).(\gamma_{B}\,\tau_{B}\,\textbf{\textsc{v}}_{B}-\gamma_{C}\,\tau_{C}\,\textbf{\textsc{v}}_{C})$.

\subsection{Eternal switching}\label{subsec:radiative-vel-et}

We proceed to evaluate $F_{AA}(\omega)$, more specifically $R_{AA}(\omega)$ denoting the transition probability rate for eternal switching, i.e., $\kappa(\tau)=1$. Here one can take help from Eq. \eqref{eq:SP-Rjj-eternal} of static probes and obtain the same result as probe $A$ is at rest. Let us now evaluate $F_{BB}(\omega)$ and $F_{CC}(\omega)$, which should have similar expressions as the Wightman functions have similar expressions in these scenarios. For instance, we consider a change of variables $s_{B}=\tau_{B}+\tau'_{B}$ and $u_{B}=\tau_{B}-\tau'_{B}$ and proceed to evaluate $F_{BB}(\omega)$,
\begin{widetext}
\begin{eqnarray}\label{eq:UVP-Fbb-eternal-1}
    F_{BB}(\omega) &=& -\frac{1}{8\pi^2}\,\int_{-\infty}^{\infty} ds_{B}\int_{-\infty}^{\infty} du_{B}\,\frac{e^{-i\,\omega\,u_{B}}}{\{\gamma_{B}(1+\textsc{v}_{B})u_{B}-i\,\epsilon\}~\{\gamma_{B}(1-\textsc{v}_{B})u_{B}-i\,\epsilon\}}~.
\end{eqnarray}
\end{widetext}
The above integrand has two pole both in the upper half complex plane. Therefore, the above integral is non-vanishing only when $\omega<0$. The final result after taking the residues carefully, see Appendix \ref{Appn:FBB-Inertial-Eternal}, will be
\begin{eqnarray}\label{eq:UVP-Fbb-eternal-2}
    F_{BB}(\omega) &=& -\frac{\omega\,\Theta(-\omega)}{4\pi}\,\int_{-\infty}^{\infty} ds_{B}~,
\end{eqnarray}
which motivates one to define a transition probability rate as $R_{BB}(\omega)(\omega) = -\omega\,\Theta(-\omega)/(4\pi)$. The expression for $R_{CC}(\omega)$ corresponding to probe $C$ can be similarly obtained.

Now we evaluate $F_{AB}(\omega)$, which utilizes the Green's function $G_{AB}$. One can notice that from the expression of $F_{AB}(\omega)$ the expression of $F_{AC}$ can be easily estimated by replacing $B$ with $C$. We consider a change of variables $s_{AB}=\tau_{A}+\tau_{B}$ and $u_{AB}=\tau_{A}-\tau_{B}$. Then the expression of $F_{AB}$ becomes
\begin{subequations}
\begin{eqnarray}
    F_{AB}(\omega) &=& -\frac{1}{8 \pi ^2 }\,\int_{-\infty}^{\infty} ds_{AB}\int_{-\infty}^{\infty} du_{AB}\,\frac{e^{-i \,u_{AB}\,\omega}}{\Delta_{AB}(u_{AB},s_{AB})} \nn \\
    \label{eq:UVP-Fab-eternal}
\end{eqnarray}
\end{subequations}
where, 
\begin{widetext}
\begin{eqnarray}\label{eq:UVP-Fab-eternal-b}
\Delta_{AB}(u_{AB},s_{AB}) &=& \frac{1}{4}\big[-4 d_{AB}^2+4 \gamma_{B}\, d_{AB}\, \textsc{v}_{B} (u_{AB}-s_{AB}) \cos{\phi_{B}} +~\big\{(1+\gamma_{B})u_{AB}+(1-\gamma_{B})s_{AB}-2 i \epsilon \big\}^2 \nn \\
&& \quad  -\gamma_{B}^2 \textsc{v}_{B}^2 (u_{AB}-s_{AB})^2\big]~.
\end{eqnarray}
\end{widetext}
In the above expression, $\phi_{B}$ is the angle between ${\bf d}_{AB}$ and $\textbf{\textsc{v}}_{B}$. Here, the integrand is not time translationally invariant, which is evident from its dependence on $s_{AB}$. Therefore, one cannot define a general rate of transition probability in this scenario. Moreover, one can evaluate the above integral numerically and see that for arbitrary nonzero velocity $\textsc{v}_{B}$, the integral remains finite. If we proceed to evaluate the rate of this cross-transition $R_{AB}(\omega)=F_{AB}(\omega)/\tilde{T}$ following \eqref{eq:Rjl-Def-a}, we will have to divide $F_{AB}(\omega)$ with an infinite quantity as $\tilde{T}=\lim_{T\to\infty}(1/2)\int_{-T}^{T}ds$. Naturally, the cross-transition probability rate $R_{AB}(\omega)$, and in a similar manner $R_{AC}(\omega)$, will vanish when the probe $B$ (or $C$) has a nonzero velocity. Note that this situation will change if $\textsc{v}_{B}\to 0$ (or $\textsc{v}_{C}\to 0$), as one can see from Eq. \eqref{eq:UVP-Fab-eternal}, in this scenario, the integral becomes independent of $s_{AB}$ and a non-vanishing rate of transition probability can be defined. However, $\textsc{v}_{B}\to 0$ or $\textsc{v}_{C}\to 0$ corresponds to the static probe case, which is already studied in our previous section and is of no interest in the present context. Therefore, we take $\textsc{v}_{B}\neq 0$ and $\textsc{v}_{C}\neq 0$, and this means that the cross-transition probability rates $R_{AB}(\omega)$ and $R_{AC}(\omega)$ vanish.

We now evaluate $F_{BC}(\omega)$, and to evaluate it we consider the Wightman function connecting two spacetime events between probes $B$ and $C$. In particular, the expression of this integral is 
\begin{eqnarray}
    F_{BC}(\omega) = -\frac{1}{4 \pi ^2 }\int_{-\infty}^{\infty} d\tau_{B}\int_{-\infty}^{\infty} d\tau_{C} ~\frac{e^{-i \,\omega\,(\tau_{B}-\tau_{C})}}{\Delta_{BC}(\tau_{B},\tau_{C})}
    \label{eq:UVP-Fbc-eternal-1}
\end{eqnarray}
where, 
\begin{widetext}
\begin{subequations}\label{eq:UVP-Fbc-eternal-1p}
\begin{eqnarray}
    \Delta_{BC}(\tau_{B},\tau_{C}) &=& (\gamma_{B}\tau_{B}-\gamma_{C}\tau_{C}-i\,\epsilon)^2-|{\bf d}_{AB}-{\bf d}_{AC}|^2-|\gamma_{B}\,\tau_{B}\,\textbf{\textsc{v}}_{B}-\gamma_{C}\,\tau_{C}\,\textbf{\textsc{v}}_{C}|^2\nonumber\\
    ~&& - 2\,({\bf d}_{AB}-{\bf d}_{AC}).(\gamma_{B}\,\tau_{B}\,\textbf{\textsc{v}}_{B}-\gamma_{C}\,\tau_{C}\,\textbf{\textsc{v}}_{C})~,
\end{eqnarray}
\end{subequations}
\end{widetext}
which becomes time translational invariant only when $\textbf{\textsc{v}}_{B}=\textbf{\textsc{v}}_{C}$. In this particular situation $\gamma_{B}=\gamma_{C}$ and unlike the previous case we do not have to choose a zero-velocity situation, and we can define the transition probability rate. When $\textbf{\textsc{v}}_{B}=\textbf{\textsc{v}}_{C}$, we have this cross-correlation term, with a change of variables $s_{BC}=\tau_{B}+\tau_{C}$ and $u_{BC}=\tau_{B}-\tau_{C}$, given by
\begin{eqnarray}
    F_{BC}(\omega) = -\frac{1}{8\pi^2}\,\int_{-\infty}^{\infty} ds_{BC}\int_{-\infty}^{\infty} du_{BC}\, \frac{e^{-i\, \omega\, u_{BC}}}{\Delta_{BC}(u_{BC},s_{BC})}~, \nn \\
    \label{eq:UVP-Fbc-eternal-2}
\end{eqnarray}
where, 
\begin{eqnarray}
    \Delta_{BC}(u_{BC},s_{BC}) &=& (\gamma\, \,u_{BC}-i \epsilon )^2-\left(d_{BC}^2+\gamma^2\, \textsc{v}^2\, u_{BC}^2+ \r. \nn \\
    && \quad \quad \l. 2 \gamma\, \textsc{v}\, d_{BC}\, u_{BC}\, \cos{\phi_{BC}}\right)~;
    \label{eq:UVP-Fbc-eternal-2b}
\end{eqnarray}
where $|{\bf d}_{AB}-{\bf d}_{AC}|=d_{BC}$, $\textbf{\textsc{v}}_{B}=\textbf{\textsc{v}}_{C}=\textbf{\textsc{v}}$, $\gamma_{B}=\gamma_{C}=\gamma$, and $\phi_{BC}$ is the angle between $({\bf d}_{AB}-{\bf d}_{AC})={\bf d}_{BC}$ and $\textbf{\textsc{v}}$. The above integrand has two poles in the upper half complex plane. Therefore, one must have $\omega<0$ to obtain any nonzero residue. After carefully finding the residues one obtains the final expression as
\begin{eqnarray}
    F_{BC}(\omega) = -\frac{\Theta(-\omega)}{4\pi}\,\frac{e^{i \, \omega \sqrt{d_{\gamma}^2-d_{BC}^2}} \sin \left(\omega d_{\gamma BC} \right)}{d_{\gamma}}\,\int_{-\infty}^{\infty} ds_{BC}~. \nn \\
    \label{eq:UVP-Fbc-eternal-3}
\end{eqnarray}
where, $d_{\gamma}= \gamma\,d_{BC} \sqrt{1-\textsc{v}^2 \sin^2{\phi_{BC}}}$. From the above expression, one can define the rate of cross transition as $R_{BC}(\omega)=F_{BC}(\omega)/\big\{\int_{-\infty}^{\infty}ds\big\}$. When $\phi_{BC}=\pi/2$, the expression of $R_{BC}(\omega)$ reduces to $R_{BC}(\omega) = -\Theta(-\omega)\,\sin \left(\omega\,d_{BC}\right)/ (4\pi\,d_{BC})$. Therefore, the cross transition becomes independent of the $\textsc{v}$, when the probes are in uniform velocity $(\textbf{\textsc{v}})$ along a direction perpendicular to the separation between $B$ and $C$. At the same time, for $\phi_{BC}=0$, the expression of the cross transition rate becomes $R_{BC}(\omega) = -\Theta(-\omega)\,\sin \left(\omega \,\gamma\,d_{BC}\right)\,e^{i \,\gamma\,\textsc{v}\,d_{BC} \omega}/(4\pi\,\gamma\,d_{BC})~$. Contrary to the previous expression, this expression depends on the velocity $\textsc{v}$ of the $B$ and $C$ probes.

It is to be noted that one can obtain the expression of $R_{CB}(\omega)$ is a similar manner, which is the cross transition probability rate corresponding to probes $C$ and $B$. In that case the only change in $R_{CB}(\omega)$, as compared to $R_{BC}(\omega)$, arrives due to the angle $\phi_{CB}$ between $({\bf d}_{AC}-{\bf d}_{AB})={\bf d}_{CB}$ and $\textbf{\textsc{v}}$. One can notice that this angle is related to $\phi_{BC}$ through the relation $\phi_{CB}=\pi-\phi_{BC}$. Then with the help of Eq. \eqref{eq:UVP-Fbc-eternal-3} we will have the expression of $R_{CB}(\omega)$ as
\begin{eqnarray}
    R_{BC}(\omega) &=& -\frac{\Theta(-\omega)}{4\pi}\,\frac{e^{-i \, \omega \sqrt{d_{\gamma}^2-d_{BC}^2}} \sin \left(\omega d_{\gamma} \right)}{d_{\gamma BC}}~, \nn \\
    &=& R_{BC}^{*}(\omega)~.
    \label{eq:UVP-Rcb-eternal}
\end{eqnarray}
The previous expression implies that $R_{CB}(\omega)$ and $R_{BC}(\omega)$ are complex conjugate to each other. Therefore, the total transition probability rate, which is obtained from a sum over all possible $R_{jl}(\omega)$ (with $j$ and $l$ being either $A$, $B$, or $C$), remains a real quantity. Here, we would like to mention that by putting these expressions of $R_{jl}(\omega)$ in Eq. \eqref{eq:Rtotal-OmTg-gen} one can get the collective transitions probability rates $\mathcal{R}_{|\Omega^{-}_{n}\rangle\to|g\rangle}(\omega)$ in this scenario, corresponding to the transitions from $|\Omega^{-}_{n}\rangle$ to $|g\rangle$.

\subsubsection{Collective transition probability rates and their characteristics}\label{subsubsec:radiative-vel-et-discussion}

\begin{figure*}
\includegraphics[width=7.8cm]{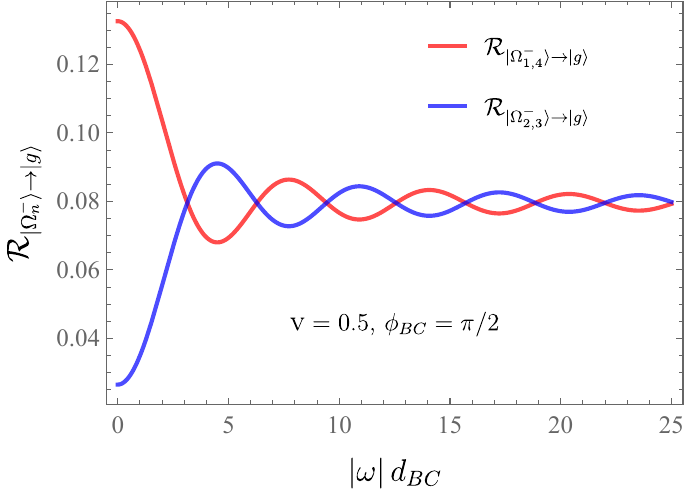}
\hskip 30pt
\includegraphics[width=7.8cm]{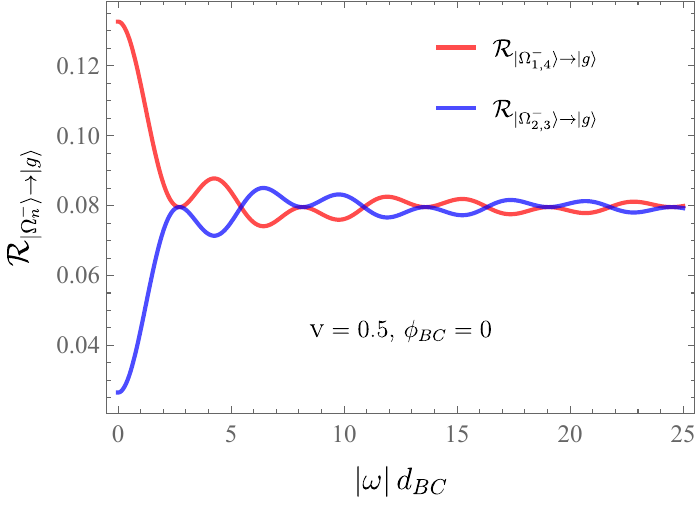}
\caption{On the left, we have plotted the collective transition probability rates $\mathcal{R}_{|\Omega^{-}_{n}\rangle\to |g\rangle}(\omega)$ as functions of the dimensionless probe energy gap $\omega\,d_{BC}$ for $\textsc{v}=0.5$ and $\phi_{BC}=\pi/2$. At the same time, on the right, the collective transition probability rates are plotted as functions of the dimensionless probe energy gap $\omega\,d_{BC}$ for $\textsc{v}=0.5$ and $\phi_{BC}=0$. It is to be noted that $\phi_{BC}$ denotes the angle between the direction of velocity $\textbf{\textsc{v}}$ of the probes ($B$ and $C$) and ${\bf d}_{BC}$. In the above plots, the red curves correspond to the transition probability rates $\mathcal{R}_{|\Omega^{-}_{1}\rangle\to |g\rangle}(\omega)$ and $\mathcal{R}_{|\Omega^{-}_{4}\rangle\to |g\rangle}(\omega)$, which happens to be equal in this scenario. While the blue curves correspond to the transition probability rates $\mathcal{R}_{|\Omega^{-}_{2}\rangle\to |g\rangle}(\omega)$ and $\mathcal{R}_{|\Omega^{-}_{3}\rangle\to |g\rangle}(\omega)$, which are the same. From both the above plots, one can notice that the different maxima or minima corresponding to the red curves appear when the blue curves have minima or maxima, respectively. However, there is a distinguishing difference between the two. For instance, for $\phi_{BC}=\pi/2$, the red and blue curves do not overlap each other at their maxima or minima. In contrast, for $\phi_{BC}=0$, the red and blue curves overlap at some of their maxima or minima.}
    \label{fig:R-vel-et-vOm}
\end{figure*}

\begin{figure*}
\includegraphics[width=7.8cm]{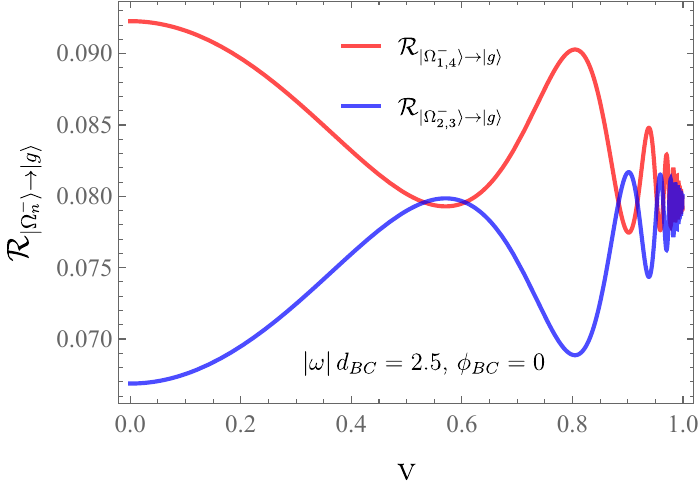}
\hskip 30pt
\includegraphics[width=7.8cm]{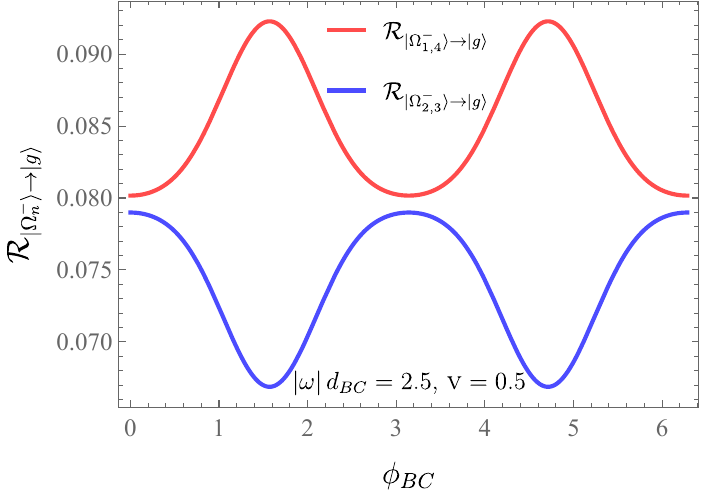}
\caption{On the left, we have plotted the collective transition probability rates $\mathcal{R}_{|\Omega^{-}_{n}\rangle\to |g\rangle}(\omega)$ as functions of the velocity $\textsc{v}$ of the probes $B$ and $C$ for fixed dimensionless probe energy gap $\omega\,d_{BC}=2.5$ and $\phi_{BC}=0$. It is to be noted that we only considered the $\phi_{BC}=0$ case as the $\phi_{BC}=\pi/2$ scenario is $\textsc{v}$ independent. On the right, we have plotted $\mathcal{R}_{|\Omega^{-}_{n}\rangle\to |g\rangle}(\omega)$ as functions of $\phi_{BC}$ for fixed dimensionless energy gap $\omega\,d_{BC}=2.5$ and $\textsc{v}=0.5$. Here also the red curves correspond to the collective transition probability rates $\mathcal{R}_{|\Omega^{-}_{1}\rangle\to |g\rangle}(\omega)$ and $\mathcal{R}_{|\Omega^{-}_{4}\rangle\to |g\rangle}(\omega)$. At the same time, the blue curves correspond to $\mathcal{R}_{|\Omega^{-}_{2}\rangle\to |g\rangle}(\omega)$ and $\mathcal{R}_{|\Omega^{-}_{3}\rangle\to |g\rangle}(\omega)$. From the above plots, one can observe that as the velocity increases and reaches ultra-relativistic, the distinct transition rates tend to converge to a specific value. At the same time, as $\phi_{BC}$ varies, we periodically get back the same features for the red and the blue curves. Specifically, we get back the same features for a $\pi$ change in the value of $\phi_{BC}$.}
    \label{fig:R-vel-et-vsv-vsPhi}
\end{figure*}

We have observed that for eternal switching and when the probes $B$ and $C$ are in uniform velocity, the only cross-correlation rates that influence the collective transition rates are the $R_{BC}$ and $R_{CB}$. With this information and with the help of Eq. \eqref{eq:Rtotal-OmTg-gen} one can get the collective transitions probability rates $\mathcal{R}_{|\Omega^{-}_{n}\rangle\to|g\rangle}(\omega)$ in this scenario, corresponding to the transitions from $|\Omega^{-}_{n}\rangle$ to $|g\rangle$ in a simplified form as
\begin{subequations}\label{eq:Rtotal-vel-et}
\begin{eqnarray}
    \mathcal{R}_{|\Omega^{-}_{1}\rangle\to |g\rangle}(\omega) &=& \mathcal{R}_{|\Omega^{-}_{4}\rangle\to |g\rangle}(\omega) 
    = \frac{1}{3}\,\bigg[\sum_{j=A}^{C}R_{jj}+ R_{\underleftrightarrow{BC}}\bigg]~, \nn \\
    \\
    \mathcal{R}_{|\Omega^{-}_{2}\rangle\to |g\rangle}(\omega) &=&
    \mathcal{R}_{|\Omega^{-}_{3}\rangle\to |g\rangle}(\omega) = \frac{1}{3}\,\bigg[\sum_{j=A}^{C}R_{jj} - R_{\underleftrightarrow{BC}}\bigg]~. \nn \\
\end{eqnarray}
\end{subequations}
From the above expressions, it is evident that in this scenario $\mathcal{R}_{|\Omega^{-}_{1}\rangle\to |g\rangle}(\omega)$ and $\mathcal{R}_{|\Omega^{-}_{4}\rangle\to |g\rangle}(\omega)$ are equal and similarly, $\mathcal{R}_{|\Omega^{-}_{2}\rangle\to |g\rangle}(\omega)$ and $\mathcal{R}_{|\Omega^{-}_{3}\rangle\to |g\rangle}(\omega)$ are equal. This is depicted in the plots for the collective transition probability rates in Figs. \ref{fig:R-vel-et-vOm} and \ref{fig:R-vel-et-vsv-vsPhi}.

In particular, in Fig. \ref{fig:R-vel-et-vOm} the collective transition probability rates $\mathcal{R}_{|\Omega^{-}_{n}\rangle\to |g\rangle}(\omega)$ are plotted as functions of the dimensionless energy gap $|\omega|\,d_{BC}$ for different values of $\phi_{BC}$ on left and right. From these plots, we observed that as the frequency $\omega$ or the separation $d_{BC}$ between the probes $B$ and $C$ increases, the red and blue curves tend to reach the same fixed value. We also observed that in both scenarios, the different maxima or minima corresponding to the red curves appear when the blue curves have minima or maxima. However, the left and right scenarios have a certain distinction in their features. Specifically, on the left for $\phi_{BC}=\pi/2$, the red and blue curves do not overlap each other at their maxima or minima. In contrast, on the right for $\phi_{BC}=0$, the red and blue curves overlap each other at some of their maxima or minima.

In Fig. \ref{fig:R-vel-et-vsv-vsPhi}, we plotted the collective transition probability rates a functions of the velocity $\textsc{v}$ and angle $\phi_{BC}$. From the plot for velocity, it is clear that as the velocity becomes ultra-relativistic, the transition rates converge to a specific value. At the same time, we observed periodic features in the collective transition probability rates as $\phi_{BC}$ varies. In particular, with a change of $\pi$ in the value of $\phi_{BC}$, one can get periodic features in the transition probability rates.

\subsection{Finite Gaussian switching}\label{subsec:radiative-vel-gsn}

Next, we consider a finite Gaussian switching function $\kappa(\tau_{j})=e^{-\tau_{j}^2/T^2}$ to estimate the transition rate of the entangled system with probes $B$ and $C$ in uniform velocities. The evaluation of the auto transition probability rate for the probe $A$ should be the same as provided in Eq. \eqref{eq:SP-Rjj-Gaussian} as this probe remains static. On the other hand, let us explicitly evaluate the quantities $F_{BB}(\omega)$, $F_{AB}(\omega)$, and $F_{BC}(\omega)$, which in turn will also provide the quantities $F_{CC}(\omega)$, $F_{AC}(\omega)$, and $F_{CB}(\omega)$.

First, we consider the evaluation of $F_{BB}(\omega)$. For the explicit evaluation of $F_{BB}(\omega)$ see Appendix \ref{Appn:FBB-Inertial-Gaussian}, and after evaluation its expression is
\begin{eqnarray}\label{eq:UVP-Fbb-gaussian}
    F_{BB}(\omega) &=& \frac{1}{4\pi}\, \bigg[e^{-\omega^2 T^2/2}-T\,\omega\sqrt{\frac{\pi}{2}}\,\erfc\left( \frac{T\,\omega}{\sqrt{2}}\right)\bigg]~,
\end{eqnarray}
which is the same as the one obtained without velocity. Therefore, whether static or with uniform velocity, single probe response, i.e., auto-correlation, remains to be the same. In contrast, the cross-correlation can distinguish the effects due to uniform velocity as compared to a static probe.

Second, let us focus on the evaluation of $F_{AB}(\omega)$. This cross-correlation term can be expressed using a change of variables $s_{AB}= \tau_{A}+ \tau_{B}$ and $u_{AB}=\tau_{A}-\tau_{B}$. Moreover, we utilize the expression of the Fourier transform $e^{-u_{AB}^2/(2 \,T^2)} = (T/\sqrt{2 \pi }) \int_{-\infty}^{\infty}d\xi \,e^{i\, \xi  u_{AB}-\xi ^2 T^2/2}$, and express $F_{AB}$ as
\begin{widetext}
\begin{eqnarray}
    F_{AB}(\omega) &=& -\frac{T}{8\pi^2 \sqrt{2\pi}}\int_{-\infty}^{\infty}ds_{AB} \int_{-\infty}^{\infty} d\xi \int_{-\infty}^{\infty} du_{AB}\, \frac{e^{i (\xi-\omega) u_{AB}
    -(\xi ^2 T^4+s_{AB}^2)/(2 T^2)}}{\Delta_{AB}(u_{AB},s_{AB})}~.
    \label{eq:UVP-Fab-gaussian}
\end{eqnarray}
\end{widetext}
The expression of $\Delta_{AB}(u_{AB},s_{AB})$ can be obtained from Eq. \eqref{eq:UVP-Fab-eternal-b}. In Eq. \eqref{eq:UVP-Fab-gaussian}, the $u_{AB}$ and then the $\xi$ integrations can be analytically carried out. After that, one can perform the $s_{AB}$ integration numerically to obtain the final result, which is finite. We would like to mention that though we have the expression of $F_{AB}(\omega)$ from Eq. \eqref{eq:UVP-Fab-gaussian} after the $u_{AB}$ and $\xi$ integrations, we will not express it explicitly in our manuscript as it is too cumbersome. Here, we would also like to mention that we have checked its expression in the limit of $\textsc{v}_{B}\to 0$, which results in the expression given by Eq. \eqref{eq:SP-Fjl-Gaussian-3} corresponding to the static probe scenario.

Next, to evaluate the cross-correlation $F_{BC}(\omega)$ we consider the specific scenario of $\textbf{\textsc{v}}_{B}=\textbf{\textsc{v}}_{C}$, which simplifies the calculations to some extent and also helps us compare the results with the eternal switching scenario. Using a change of variables $s_{BC}= \tau_{B} +\tau_{C}$ and $u_{BC}=\tau_{B}-\tau_{C}$ and the Fourier transform of the Gaussian factor, we express
\begin{widetext}
\begin{eqnarray}\label{eq:UVP-Fbc-Gaussian-1}
    F_{BC}(\omega) &=& -\frac{T}{8\pi^2\sqrt{2\pi}}\,\int_{-\infty}^{\infty} ds_{BC}\int_{-\infty}^{\infty} d\xi \int_{-\infty}^{\infty} du_{BC}\, \frac{e^{i\, (\xi-\omega)\, u_{BC}}~e^{-(\xi ^2 T^4+s_{BC}^2)/(2 T^2)}}{\Delta_{BC}(u_{BC},s_{BC})}~,
\end{eqnarray}
\end{widetext}
where the expression of $\Delta_{BC}(u_{BC},s_{BC})$ can be obtained from Eq. \eqref{eq:UVP-Fbc-eternal-2b}. Here also the procedure to obtain the final expression is to integrate over $u_{BC}$ first. Then one will have to carefully carry out the integrations over $\xi$ and $s_{BC}$ respectively. Unlike the evaluation of $F_{AB}$, in this scenario, we could obtain an exact analytical expression for $F_{BC}$. Although this expression is a bit cumbersome to look at, and we have relegated it to the Appendix, see Eq. \eqref{eq:UVP-Fcb-Gaussian-2} in Appendix \ref{Appn:FBC-Inertial-Gaussian}.\vspace{0.2cm}

\begin{figure*}
\includegraphics[width=7.8cm]{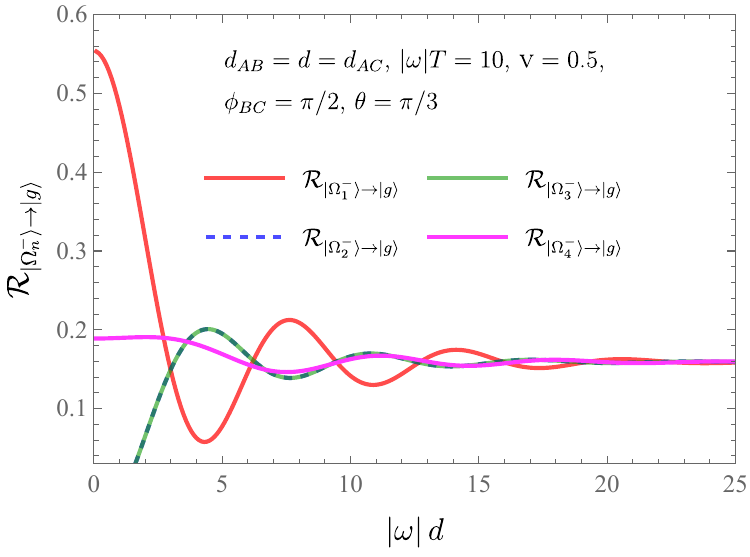}
\hskip 30pt
\includegraphics[width=7.8cm]{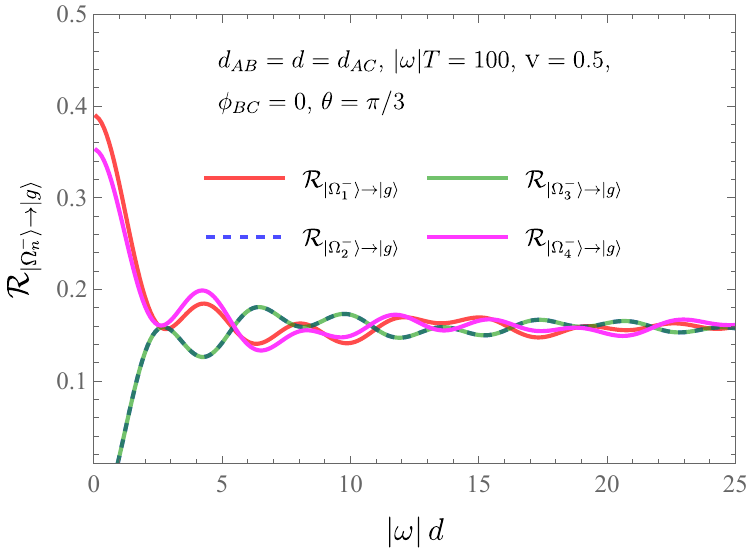}
\vskip 10 pt
    \includegraphics[width=7.8cm]{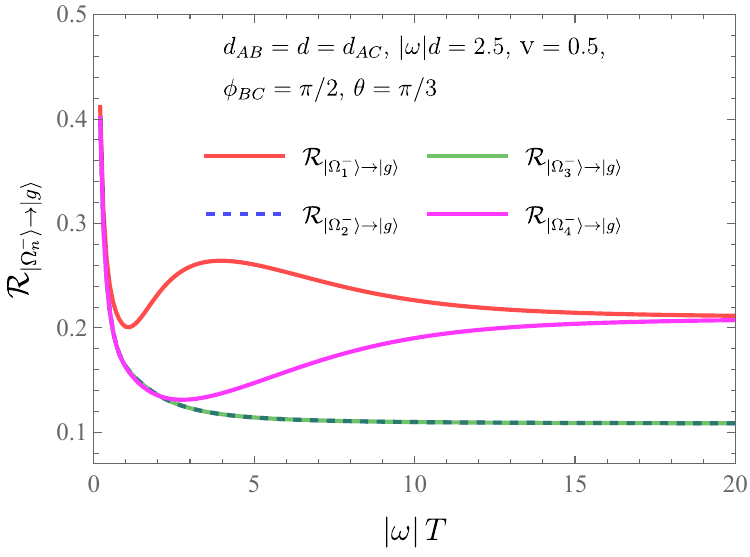}
\hskip 30pt
\includegraphics[width=7.8cm]{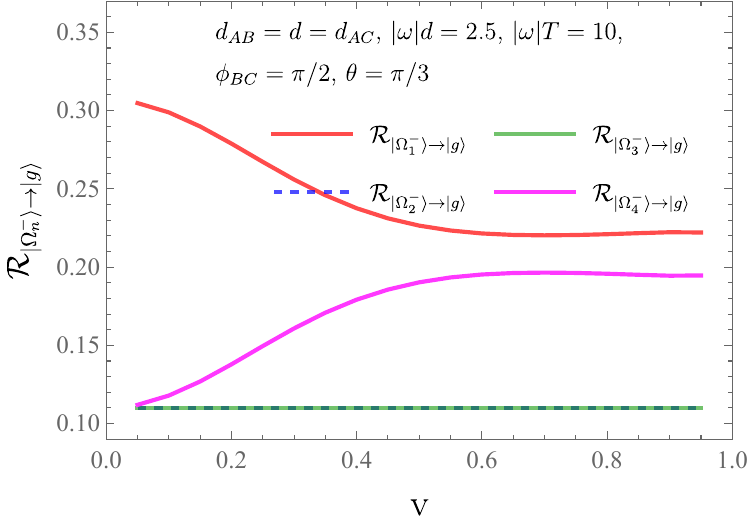}
\caption{\textbf{Top-left:} We have plotted the collective transition probability rates $\mathcal{R}_{|\Omega^{-}_{n}\rangle\to|g\rangle}(\omega)$ as functions of the dimensionless probe energy gap $|\omega|d$ for a scenario where the three probes initial positions are at the three vertices of an equilateral triangle and for Gaussian switching. We have considered a scenario where $d_{AB}=d=d_{AC}$, $\theta=\pi/3$, $\textsc{v}=0.5$, $\phi_{BC}=\pi/2$, and $|\omega|T=10$. \textbf{Top-right:} We have plotted the collective transition probability rates as functions of the dimensionless probe energy gap $|\omega|d$ for a scenario where the three probes' initial positions are at the three vertices of an equilateral triangle and for Gaussian switching. We have considered a scenario where $d_{AB}=d=d_{AC}$, $\theta=\pi/3$, $\textsc{v}=0.5$, $\phi_{BC}=0$, and $|\omega|T=100$. From the top two plots, one can observe that smaller switching times lead to fewer ripples in the transition probability rates. We also observe that the transitions from the states $|\Omega^{-}_{2}\rangle$ and $|\Omega^{-}_{3}\rangle$ are equal in all scenarios, which is similar to the eternal switching scenario, see Fig. \ref{fig:R-vel-et-vOm}. At the same time, contrary to the eternal switching scenario, the transitions from the states $|\Omega^{-}_{1}\rangle$ and $|\Omega^{-}_{4}\rangle$ are different here. However, when the switching time increases, see the top-right plot, the transitions from the states $|\Omega^{-}_{1}\rangle$ and $|\Omega^{-}_{4}\rangle$ tend to approach similar values. \textbf{Bottom-left:} The collective transition probability rates are plotted as functions of the dimensionless switching time $|\omega|T$ for $|\omega|d = 2.5$, $\theta=\pi/3$, $\textsc{v}=0.5$, and $\phi_{BC}=\pi/2$. From this plot also one can observe that as $T$ increase $(|\omega|\,T\sim20)$, the transition probability rates from the states $|\Omega^{-}_{1}\rangle$ and $|\Omega^{-}_{4}\rangle$ reach similar values. \textbf{Bottom-right:} The collective transition probability rates are plotted as functions of the velocity $\textsc{v}$ for $|\omega|d=2.5$, $|\omega|T=10$, $\theta=\pi/3$, and $\phi_{BC}=\pi/2$. From the bottom-left plot, we observe that for very low switching time $(|\omega|T<1)$ all the transition probability rates increase with decreasing $T$, and for $|\omega|T\ge 2$ this feature is not the same for all the probability rates. These features are similar to the static probe scenario. All of these above plots in this figure are obtained for Gaussian switching of $\kappa(\tau_{j})=e^{-\tau_{j}^2/T^2}$.}
    \label{fig:R-vel-Gsn-vOm}
\end{figure*}

\subsubsection{Collective transition probability rates and their characteristics}\label{subsubsec:radiative-vel-gsn-discussion}

In this part of the section, we discuss characteristics of the collective transition probability rates $\mathcal{R}_{|\Omega^{-}_{n}\rangle\to |g\rangle}(\omega)$ corresponding to probes $B$ and $C$ in uniform velocity and for Gaussian switching function. One can obtain these collective transition probability rates by putting the values of $F_{jl}(\omega)$ as obtained numerically from Eqs \eqref{eq:UVP-Fbb-gaussian}, \eqref{eq:UVP-Fab-gaussian}, \eqref{eq:UVP-Fbc-Gaussian-1} and using Eqs. \eqref{eq:Rjl-Def} and \eqref{eq:Rtotal-OmTg-gen}. To understand the effects of finite window function as compared to the eternal switching, we will consider parameter values that facilitate the comparison with Fig. \ref{fig:R-vel-et-vOm}. In this regard, we have considered $d_{AB}=d_{AC}=d$ and $\theta=\pi/3$, i.e., the positions of the probes at $\tau=0$ are at the vertices of an equilateral triangle. We have also considered $\phi_{B}=\pi/2$, $\phi_{C}=\pi/2$, and $\textbf{\textsc{v}}_{B}=\textbf{\textsc{v}}_{C}=\textbf{\textsc{v}}$. With the help of the transition coefficients from Eqs. \eqref{eq:UVP-Fbb-gaussian}, \eqref{eq:UVP-Fab-gaussian}, and \eqref{eq:UVP-Fbc-Gaussian-1}, and the general expression for the collective transition probability rates \eqref{eq:Rtotal-OmTg-gen}, one can obtain the transition probability rates for transitions from $|\Omega^{-}_{n}\rangle$ to $|g\rangle$.

In Fig. \ref{fig:R-vel-Gsn-vOm}, top-panel, we plotted these collective transition probability rates as functions of the dimensionless energy gap $|\omega|d$ for $\phi_{BC}=\pi/2$ and $\phi_{BC}=0$, respectively. These plots resemble the ones with eternal switching, such as the transitions from $|\Omega^{-}_{2}\rangle$ and $|\Omega^{-}_{3}\rangle$ are equal. However, contrary to eternal switching the transitions from $|\Omega^{-}_{1}\rangle$ and $|\Omega^{-}_{4}\rangle$ are not the same, though they tend to reach similar values if the switching time is made large. In Fig. \ref{fig:R-vel-Gsn-vOm}, in bottom-panel, we have plotted the collective transition probability rates as functions of the dimensionless switching time $|\omega|T$ and velocity $\textsc{v}$ respectively. From the bottom-left plot also one can observe that for large switching time $(|\omega|T\sim 20)$, the transitions from the states become quantitatively and qualitatively similar.

\section{Effect of thermal bath in tripartite radiative process}\label{sec:thermal-bath}

In this section, we analyze the effect of a thermal bath on the transition rates of quantum probes prepared in a tripartite-entangled state. As discussed earlier, determining both the single and collective transition rates involves evaluating the pullback of the Wightman function of the quantum field. So far, we have used the vacuum expectation value of the field correlators to compute the Wightman two-point function. However, in the presence of a thermal bath, the field is in a thermal state, and the appropriate thermal Wightman function must be used to account for the radiative processes contributing to the transition rates. In $(n+1)$-dimensional flat spacetime, the Green function in a thermal state at temperature $T=1/\beta$ is given by \cite{Kolekar:2013hra, Kolekar:2013xua},
\begin{align}
    G^{+}_{\beta}(x,x^\prime) = \frac{1}{(2\pi)^n}\int \d^n k & \l[\f{e^{-i\omega_k(t-t^\prime)+i{\bf k}({\bf x} - {\bf x}^\prime)}}{2\omega_k(1-e^{-\beta \omega_k})} \r. \nn \\
    & \l. + \f{e^{-i\omega_k(t-t^\prime)+i{\bf k}({\bf x} - {\bf x}^\prime)}}{2\omega_k(e^{\beta \omega_k}-1)} \r]~.
\end{align}
For our convenience, we use the spacetime form of the Wightman function,
\begin{widetext}
\begin{align}
    G^{+}_{\beta}\l(x(\tau),x^\prime(\tau)\r) = -\f{1}{4\pi^2}\sum_{n=-\infty}^{\infty} \f{1}{\l\{\Delta t(\tau, \tau^\prime)-i n \beta - i \epsilon\r\}
    ^2 - \l\{\Delta {\bf x} (\tau,\tau^\prime)\r\}^2}~.
\end{align}
\end{widetext}
where $\Delta t(\tau, \tau^\prime) = t(\tau)-t'(\tau^\prime) $ and $\Delta {\bf x}(\tau,\tau^\prime) = |{\bf x}(\tau)-{\bf x}'(\tau')| =\l(x(\tau)-x(\tau^\prime)\r)^2 + \l(y(\tau)-y(\tau^\prime)\r)^2 + \l(z(\tau)-z(\tau^\prime)^2 \r)$ and $\tau$ is the proper time along the considered trajectory. This Wightman function is periodic in Euclidean time $(\tau \to i\tilde{\tau} + 2n\pi \beta)$, which is one of the conditions that needs to be satisfied by the Wightman function to be thermal. It should be noted that this condition alone may not sufficient for a state to be thermal state. One such example was demonstrated in \cite{K:2023dmj}. 

Since the field is not in the vacuum state, but rather in the thermal state, the probes can de-excite as well as excite. When we consider the probes initially prepared in the entangled state, $\ket{\Omega^{-}_{n}}$, due to the thermal bath, the collective state of the tripartite system can de-excite from $\ket{\Omega^{-}_{n}}$ to ground state and also can excite from $\ket{\Omega^{-}_{n}}$ to $\ket{\Omega^{+}_{n}}$. We will analyze both cases in this section. 

Note that from the discussions on radiative process with the switching function, it is clear that the only effect of the smooth switching is that it dampens the collective response of the probes and diminishes the distinguishability of each transition, keeping the overall features of the response function intact. Hence in the upcoming discussion for the thermal bath we will focus on the eternal switching and the features here should propagate to case when switching functions are considered.

\subsection{Static probes}
Here we consider the scenario with static quantum probes. We start by evaluating the individual transition probability rates of the probes. Then we will evaluate the cross-transition probability rates and finally obtain the collective transition probability rates.

\subsubsection{Individual transition probability rates}
When the probes are at rest inside a thermal bath, their respective trajectories are same as in Eq. \eqref{eq:SP-trajectories}. A schematic diagram depicting these static probes is also provided in Fig. \ref{fig:static-detectors}. Substituting the trajectory to find the pullback of the thermal Wightman function gives,
%

\begin{align}
    G^{+}_{(\beta)jj} = -\frac{1}{4\pi^2} \sum_{n=-\infty}^{\infty}\frac{1}{(u_j -in\beta-i\epsilon)^2}~,
\end{align}
where $u_{j}=\tau_j-\tau_{j}^\prime$. Even though this series has an infinite term sum, this series can be exactly summed in terms of hyperbolic functions\cite {book:Birrell}. Now, the individual response rate can be found from the definition used in Eq. \ref{eqn:ind-rate-eternal} as,
\begin{align}
    R_{jj}(\omega) = -\frac{1}{4\pi^2} \int_{-\infty}^{\infty} \d u_j \,e^{-i\omega u_j} \sum_{n=-\infty}^{\infty}\frac{1}{(u_j-i n\beta-i \epsilon)^2}~.
\end{align}
This integration can be done in a straightforward manner using contour integration. By analytically continuing $u$, we can see that the poles are at $u=i\pi\epsilon+i n \beta$. For the de-excitation, $\omega<0$, the contour is closed in the upper half-plane, which gives the transition rate as,
\begin{align}
    R_{jj}^{\rm em}(\omega) = -\frac{\omega}{2\pi}\l[ \f{\Theta(-\omega)}{1-e^{\beta \omega}}\r]~.
    \label{eq:TH-ind-de-excite}
\end{align}
For the excitation from the first excited state, $\ket{\Omega^{-}_{n}}$ to second excited state $\ket{\Omega^{+}_{n}}$, with the same energy gap, $\omega>0$, the rate of excitation can be obtained by closing the contour in the lower half-plane as,
\begin{align}
    R_{jj}^{\rm abs}(\omega) = \f{\omega}{2\pi}\l( \f{1}{e^{\beta \omega}-1}\r)~.
    \label{eq:TH-ind-excite}
\end{align}
The response rates for spontaneous emission and absorption obey the detailed balance condition, $\mathcal{R}_{jj}^{\rm em}/\mathcal{R}_{jj}^{\rm abs} = e^{\beta\omega}$.

\subsubsection{Collective transition probability rates and their characteristics}
To evaluate the collective probability rates, first, the cross transition probabilities are calculated, which requires the spacetime interval between each pair of probes to be used in the Wightman function. The trajectories are the same as the one taken for the case without a thermal bath (Eq. \eqref{eq:SP-trajectories}). The correct thermal Wightman function for evaluating the cross probe transition rate in the case of static probes is,
\begin{align}
    G^{+}_{(\beta)jl}=\f{1}{4\pi^2}\sum_{-\infty}^{\infty} \f{1}{-(u_{jl}-in\beta-i\epsilon)^2+d^2}\,.
\end{align}
Then the transition rates will be,
\begin{align}
    R_{jl}(\omega)=\f{1}{4\pi^2}\int_{-\infty}^{\infty}\d u_{jl} \,e^{-i\omega u_{jl}} \sum_{-\infty}^{\infty} \f{1}{-(u_{jl}-in\beta-i\epsilon)^2+d^2}\,.
\end{align}
Unlike individual probe transition rates, the poles are at $\pi d+ i\pi \epsilon+i\pi m\beta$ and $-\pi d+ i\pi \epsilon+i\pi m\beta$, which are not purely imaginary. The contour can be closed in the upper half-plane for de-excitation, $\ket{\Omega^{-}_{n}}\to \ket{g}$, and closed in the lower half-plane for excitations, $\ket{\Omega^{-}_{n}}\to \ket{\Omega^{+}_{n}}$.

Let us look at emission as well as absorption separately. The transition rate for emission is obtained by evaluating the contour integral by closing it in the upper half-plane $(\omega<0)$ and the final expression is,
\begin{align}
    R_{jl}^{\rm em}(\omega) &= -\frac{1}{2\pi\,d}\left[\frac{\Theta(-\omega)\,\sin{\l(d_{jl}\,\omega\r)}}{1-e^{\beta\omega}}\right]~, \\
    R_{jl}^{\rm abs}(\omega) &= -\frac{1}{2\pi\,d}\left[ \frac{\Theta(\omega)\,\sin{\l(d_{jl}\,\omega\r)}}{e^{\beta\omega}-1}\right]~.   
    \label{eq:coll-thermal-resp-rate}
\end{align}
Note that individual response rates pairs $(R_{jl})$ for emission and absorption will obey detailed balance condition, but the probability rates $(\mathcal{R}_{jl})$ for both absorption for the transition $\ket{\Omega^{-}_{n}}\to\ket{\Omega^{+}_{n}}$ and emission for $\ket{\Omega^{-}_{n}}\to\ket{g}$ will \textit{not obey} the detailed balance conditions, since the value for monopole moment operator for each of these transitions are different and they have to be accounted while calculating probability rates (see Eqs. \eqref{eq:transition-prob},\eqref{eq:response-fn}). 

Using the expectation values of the monopole moment operator for the transition $\ket{\Omega^{-}_{n}}\to\ket{g}$, as before, the collective transition probability rates are the same as those obtained in Eq. \eqref{eq:Rtotal-st-et} of Sec. \ref{subsec:radiative-static-eternal} but with an overall multiplicative factor of $(e^{\beta\omega}-1)^{-1}$. Now, the expectation values of the monopole moment operator for the transition $\ket{\Omega^{-}_{n}}\to\ket{\Omega^{+}_{n}}$, can be evaluated using $m_{j}^{\omega^\prime \omega^{\prime\prime}}=\langle \omega^\prime|~m_{j}(0)~|\omega^{\prime\prime}\rangle$, see Table \ref{tab:monopole-moment}.
%

\begin{table}[h!]
    \centering
    \renewcommand{\arraystretch}{1.3}
\begin{tabular}{|c|c|c|c|c|}
    \hline
    \multicolumn{5}{|c|}{List of monopole moment expectation values} \\
    \hline
     & $\ket{\Omega^{-}_{1}}$ & $\ket{\Omega^{-}_{2}}$ & $\ket{\Omega^{-}_{3}}$ & $\ket{\Omega^{-}_{4}}$ \\
    \hline
    $\bra{\Omega^{+}_{1}}$ & $\frac{2}{3}(1,1,1)$ & $\frac{2}{3}(0,1,0)$ & $\frac{2}{3}(1,0,0)$ & $\frac{2}{3}(0,0,-1)$ \\
    \hline
    $\bra{\Omega^{+}_{2}}$ & $\frac{2}{3}(0,1,0)$ & $\frac{2}{3}(-1,1,-1)$ & $\frac{2}{3}(0,0,-1)$ & $\frac{2}{3}(-1,0,0)$ \\
    \hline
    $\bra{\Omega^{+}_{3}}$ & $\frac{2}{3}(1,0,0)$ & $\frac{2}{3}(0,0,1)$ & $\frac{2}{3}(1,-1,-1)$ & $\frac{2}{3}(0,-1,0)$ \\
    \hline
    $\bra{\Omega^{+}_{4}}$ & $\frac{2}{3}(0,0,-1)$ & $\frac{2}{3}(-1,0,0)$ & $\frac{2}{3}(0,-1,0)$ & $\frac{2}{3}(-1,-1,1)$ \\
    \hline
\end{tabular}
\caption{In the above table, we provide the expectation values of the monopole moment operators corresponding to different probes between the states $\ket{\Omega^{+}_{n}}$ and $\ket{\Omega^{-}_{n}}$. In particular, the three numbers for the expectation values inside the parentheses correspond to the monopole moment operators of different probes, namely probe $A$, $B$, and $C$.}
\label{tab:monopole-moment}
\end{table}
Using the values of the monopole moment and assuming that the three probes are exactly identical, the collective probability rates for transition between different degenerate states of $\ket{\Omega^{-}}$ and $\ket{\Omega^{+}}$ are:
\begin{subequations}
\begin{align}
    \mathcal{R}_{\ket{\Omega^{-}_{n}}\to \ket{\Omega^{+}_{n}}}(\omega) &= \frac{4}{9}\Big[ \sum_{j=A}^{C} R_{jj} + (-1)^{b_1} R_{\underleftrightarrow{AB}} + (-1)^{b_2} R_{\underleftrightarrow{AC}} \nn \\
    & \quad + (-1)^{b_1+b_2} R_{\underleftrightarrow{BC}} \Big]~,\\
    \mathcal{R}_{\ket{\Omega^{-}_{n}}\to \ket{\Omega^{+}_{m}}}(\omega) &= \frac{4}{9} R_{jj}^{\rm abs}(\omega)~, \quad {\rm for }\; n\ne m~.
\end{align}
\end{subequations}
where, $R_{\underleftrightarrow{jl}} = R_{jl} + R_{lj}$ and $b_1$, $b_2$ are the same definitions as given in Eq. \eqref{eq:binary-decomposition}. Since the collective probability transition rate for the case $n\ne m$ is equal to the individual response rate, it does not depend on the distance between each probe. Also, the magnitude of the transition probability rate for each individual transition is small compared to the collective transition probability rate. Hence, out of 16 transitions that are possible between $\ket{\Omega^{-}_{n}}$ and $\ket{\Omega^{+}_{m}}$, the dominant ones are for $n=m$ and for $n\ne m$ they are subdominant. Therefore, we define an effective transition rate which encapsulates the transition, $\ket{\Omega^{-}_{n}}\to \sum_{m=1}^{4}\ket{\Omega^{+}_m}$ for each $n$. The overall behavior of this quantity will be predominantly determined by the collective transition probability rate with $n=m$.
%
\begin{figure*}[t!]
\includegraphics[width=7.8cm]{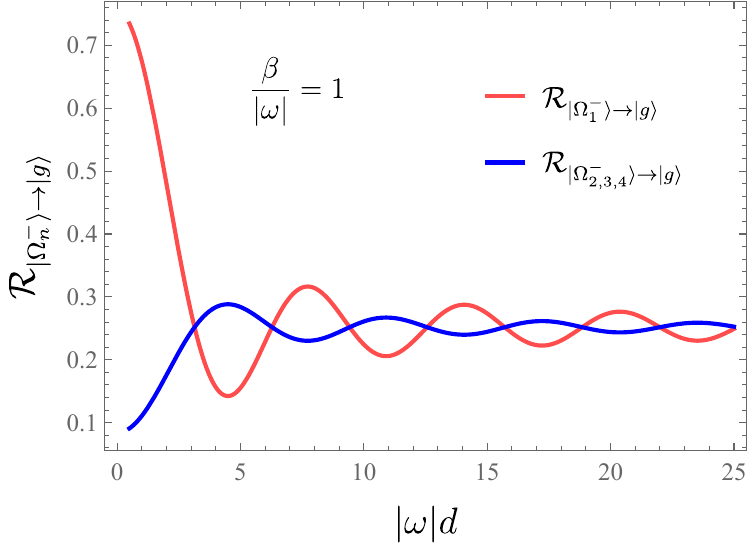}
\hskip 30pt
\includegraphics[width=7.8cm]{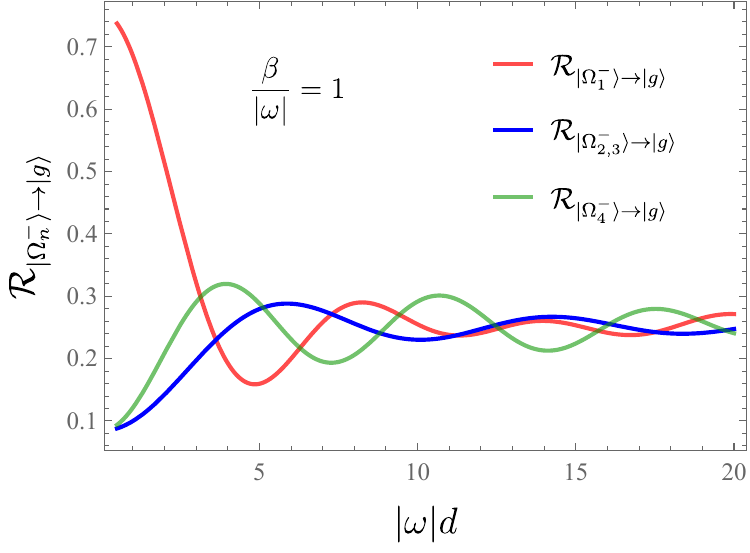}
\caption{The thermal bath leads to an increase in the collective transition probability rate for de-excitation. As the temperature increases, $\beta\to0$, the transition rates further increase due to the overall Planckian factor in the transition probability rate. All other qualitative features are similar to those of the zero temperature case. \textbf{Left}: Equilateral triangle configuration for the probes and \textbf{right}: Isosceles triangle configuration.}
    \label{fig:R-vs-wd-thermal1}
\end{figure*}
%
\begin{figure*}[t!]
\includegraphics[width=7.8cm]{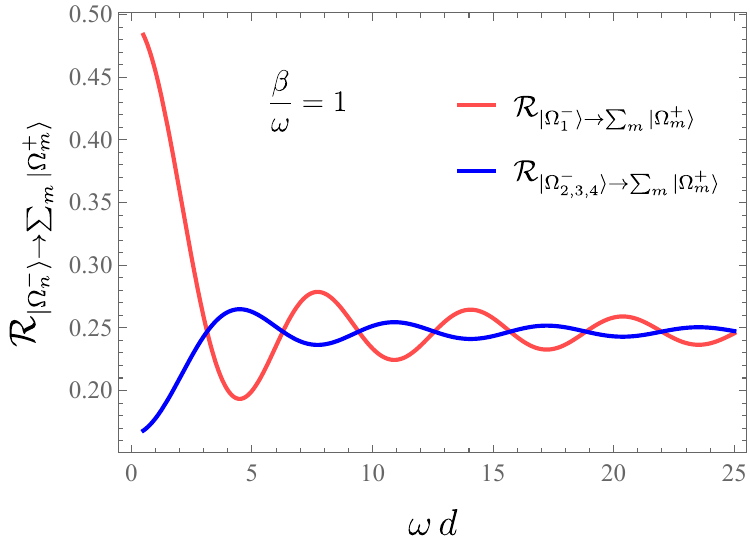}
\hskip 30pt
\includegraphics[width=7.8cm]{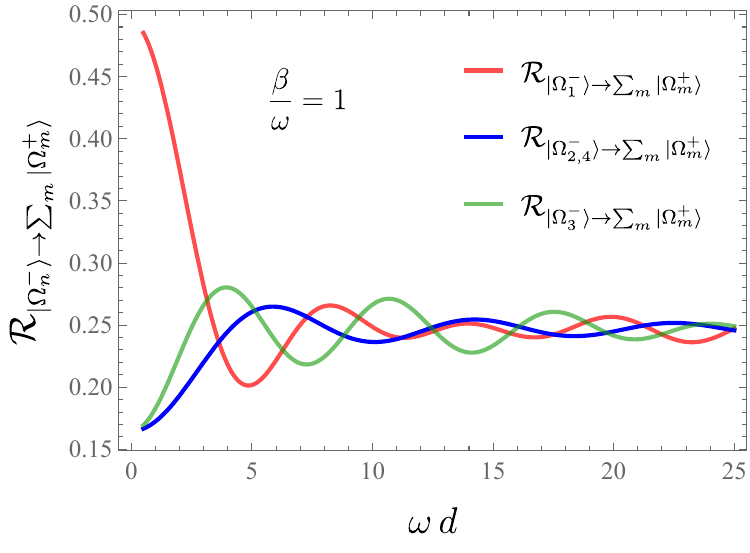}
\caption{The thermal bath leads to excitation of probes, which was not present when the field was in the vacuum state. The magnitude of the collective transition probability rate for excitation is less than that for de-excitation. The qualitative features are similar to those of the de-excitations. \textbf{Left}: Equilateral triangle configuration for the probes and \textbf{Right}: Isosceles triangle configuration.}
    \label{fig:R-vs-wd-thermal2}
\end{figure*}
%

In Fig. \ref{fig:R-vs-wd-thermal1}, we have plotted the collective transition probability rates of the probes for de-excitation from $\ket{\Omega^{-}_{n}}$ to $\ket{g}$. For demonstration, we have only plotted for the equilateral and isosceles triangle configurations, but the enhancement of transition probability rates will be the same for other configurations as well. In the case of excitation due to the presence of a thermal bath, which was not previously possible, \textit{the collective transition probability rate is lower than that of excitation} (see Fig. \ref{fig:R-vs-wd-thermal2}).  

\begin{figure*}[h!]
\includegraphics[width=7.8cm]{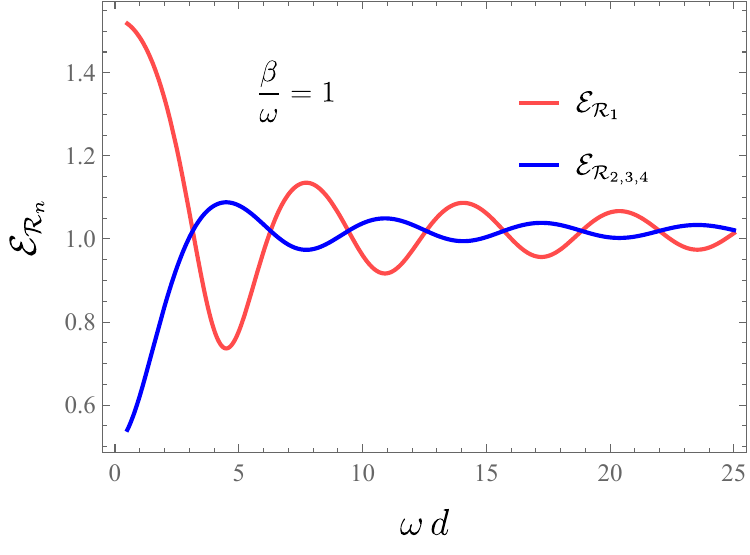}
\hskip 30pt
\includegraphics[width=7.8cm]{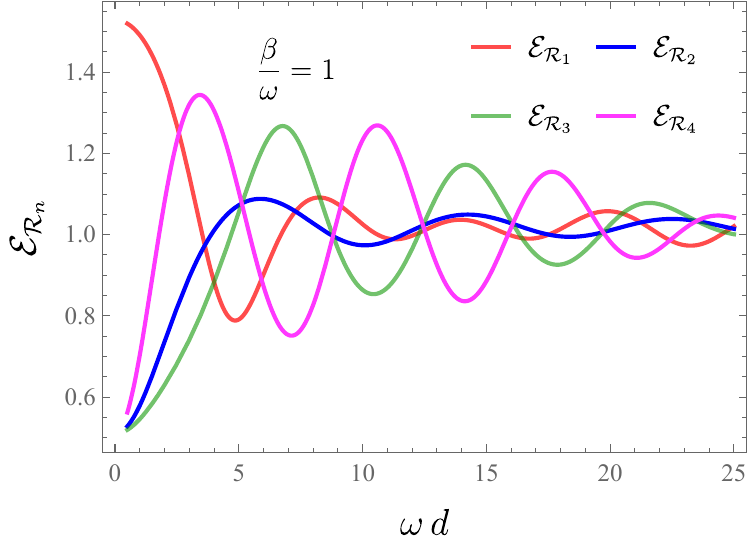}
\caption{The ratio of de-excitation and excitation transition rates with respect to the scaled distance, $|\omega| d$, between the probes by keeping $\beta/|\omega|$ constant. \textbf{Left}: Equilateral triangle configuration for the probes where the curves for $n=2,3,4$ are indistinguishable like the collective transition probability rates. \textbf{Right}: Isosceles triangle configuration where the curves are distinguishable for all the transitions, unlike the collective transition probability rates.}
    \label{fig:E-vs-wd-thermal}
\end{figure*}

We also define the quantity $\mathcal{E}_{\mathcal{R}}$ as the ratio of transition probability rate of de-excitation to excitation, i.e, $\mathcal{E}_{\mathcal{R}_n}=\mathcal{R}_{\ket{\Omega^{-}_{n}}\to\ket{g}}/\mathcal{R}_{\ket{\Omega^{-}_{n}}\to\sum_{m}\ket{\Omega^{+}_{n}}}$. For equilateral triangle configuration, this ratio do not distinguish between transitions, $\ket{\Omega^{-}_{2,3,4}}\to\ket{g}$, which shows that it has similar properties as the transition probability rates. But interestingly, for the isosceles triangle, since the transition from $\ket{\Omega^{-}_{2,3}}\to \ket{g}$ are same and for transition from $\ket{\Omega^{-}_{2,4}} \to \sum_{m}\ket{\Omega^{+}_{n}}$ are same, this causes, a complete distinguishable curves for all $\mathcal{E}_{\mathcal{R}_{n}}$. Figure \ref{fig:E-vs-wd-thermal} depicts this difference between the trend in each collective transition probability rate and ratio of transitions.
%
%

\subsection{Probes in inertial motion}
To study the effects of a thermal bath on the radiative process of tripartite entangled two-level quantum probes with relative motion, the trajectory in Eq. \ref{eq:UVP-trajectories} is considered. Since the individual probe transition rate is invariant under all inertial motions, it will be the same as given in Eqs. \ref{eq:TH-ind-de-excite} $\&$ \ref{eq:TH-ind-excite} for de-excitation and excitation respectively. At the same time, to obtain the cross transition probability rates, one needs the Wightman function connecting two events corresponding to two different probes
\begin{align}
    G^{+}_{jl}=-\sum_{n=-\infty}^{\infty}\frac{1}{4\pi}\frac{1}{(t_j-t_l-in\beta-i\epsilon)^2-|{\bf x}_j-{\bf x}_l|^2}~.
    \label{eq:TH-inertial-intr-trans}
\end{align}
We consider a probe configuration same as the one considered in Sec. \ref{subsec:radiative-vel-et}, i.e., probe $A$ is static and probes $B$ and $C$ are in uniform velocity with $\textbf{\textsc{v}}_{B}=\textbf{\textsc{v}}_{C}$. Part of the reason behind this particular choice is that it facilitates the comparison between the thermal and non-thermal scenarios. Then, proceeding in the same route as in Sec. \ref{subsec:radiative-vel-et}, the cross probe transition probability rate $R_{AB},R_{AC}$ will vanish, since we have finite transition probabilities $F_{AB}$ and $F_{AC}$. While estimating the cross transition probability rate $R_{BC}$, the equation similar to Eq. \ref{eq:UVP-Fbc-eternal-2} will give us the corresponding cross transition probability $F_{BC}(\omega)$ as,
\begin{widetext}
\begin{eqnarray}\label{eq:TH-UVP-Fbc-eternal-2}
    F_{BC}(\omega) &=& -\frac{1}{8\pi^2}\,\sum_{n=-\infty}^{\infty}\,\int_{-\infty}^{\infty} ds_{BC}\int_{-\infty}^{\infty} du_{BC}\, \frac{e^{-i\, \omega\, u_{BC}}}{\Delta_{BC}(u_{BC},s_{BC})}~,\\\nonumber
    \label{eq:TH-UVP-Fbc-eternal-2b}
\end{eqnarray}
\end{widetext}
where, $\Delta_{BC}(u_{BC},s_{BC}) = (\gamma\, \,u_{BC}-in\beta-i \epsilon )^2-\left(d_{BC}^2+\gamma^2\, \textsc{v}^2\, u_{BC}^2+2 \gamma\, \textsc{v}\, d_{BC}\, u_{BC}\, \cos{\phi_{BC}}\right)$. The integral over $u_{BC}$ in the above expression can be performed utilizing the residue theorem of complex analysis, in a manner similar to the zero temperature scenario. At the same time, the integrand is independent of $s_{BC}$, and by dividing the integral by $\tilde{T}=\lim_{T\to\infty}(1/2)\int_{-T}^{T}ds_{BC}$ we obtain the cross probe transition probability rate $R_{BC}$. One can check Eqs. \eqref{eq:RBC-Th-Vel-Oml0} and \eqref{eq:RBC-Th-Vel-Omg0} of Appendix \ref{Appn:RBC-Th-Inertial} for the expressions of $R_{BC}$, where the integration over $u_{BC}$ is carried out. However, to the best of our understanding, the sum over $n$ cannot be performed analytically, and we used numerical methods to obtain this sum. Here, we would like to mention that the cross-transition probability rate $R_{CB}$ can be obtained from $R_{BC}$ with the understanding that $\mathbf{d}_{CB}=-\mathbf{d}_{BC}$.

\begin{figure*}
\includegraphics[width=7.8cm]{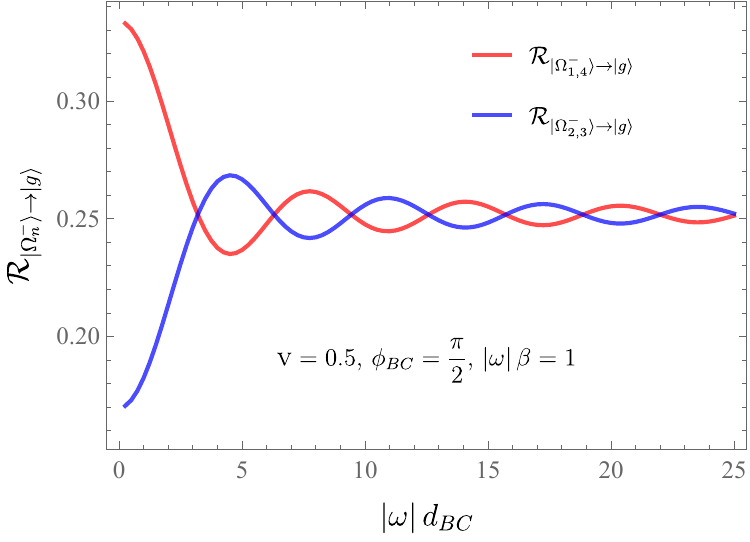}
\hskip 30pt
\includegraphics[width=7.8cm]{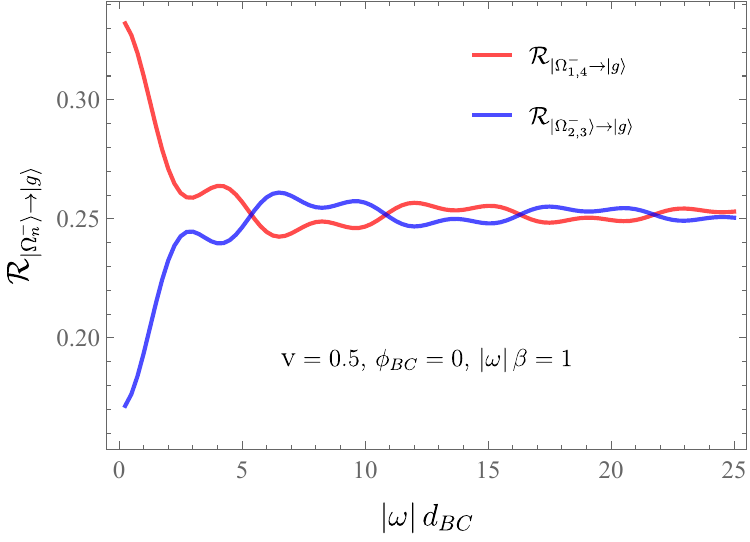}
\caption{On the left, we have plotted the collective transition probability rates $\mathcal{R}_{|\Omega^{-}_{n}\rangle\to |g\rangle}(\omega)$ as functions of the dimensionless probe energy gap $\omega\,d_{BC}$ for $\textsc{v}=0.5$, $|\omega|\beta=1$, and $\phi_{BC}=\pi/2$. At the same time, on the right, the collective transition probability rates are plotted as functions of the dimensionless probe energy gap $\omega\,d_{BC}$ for $\textsc{v}=0.5$, $|\omega|\beta=1$, and $\phi_{BC}=0$. $\phi_{BC}$ denotes the angle between the direction of velocity $\textbf{\textsc{v}}$ of the probes ($B$ and $C$) and ${\bf d}_{BC}$. Similar to the non-thermal scenario, for both the above plots, the different maxima or minima corresponding to the red curves appear when the blue curves have minima or maxima, respectively. However, compared to the non-thermal scenario, all the transition probability rates have larger quantitative values inside a thermal bath. Moreover, here, for $\phi_{BC}=0$, the red curve does not touch the blue curve at its different maxima and minima, though this situation may change for low enough temperature, such as for $|\omega|\beta=10$.}
    \label{fig:RTh-vel-et-vOm}
\end{figure*}

\subsubsection{Collective transition probability rates and their characteristics}

In this part, we investigate the collective transition probability rates. Similar to the zero temperature scenario, here also, only the cross-transition probability rates $R_{BC}$ and $R_{CB}$ remain non-vanishing, resulting from the initial consideration of $\textbf{\textsc{v}}_{B}=\textbf{\textsc{v}}_{C}$. The representation of the collective transition probability rates in terms of the auto and cross-transition probabilities is the same as provided in Eq. \eqref{eq:Rtotal-vel-et}. In Fig. \ref{fig:RTh-vel-et-vOm} we have plotted these collective transition probability rates as functions of the dimensionless parameter $\omega\,d_{BC}$ for two different values of $\phi_{BC}$. In particular, on the left, we have plotted the collective transition probability rates for $\phi_{BC}=\pi/2$. At the same time, on the right, we have plotted the probability rates for $\phi_{BC}=0$. The qualitative features of the transition probability rates from these plots should be compared to the ones from Fig. \ref{fig:R-vel-et-vOm}, corresponding to the non-thermal scenario. One can notice that compared to the non-thermal scenario, all the transition probability rates have larger quantitative values inside a thermal bath. Therefore, the presence of a thermal bath has an enhancing effect on the transition probability rates. Moreover, for $\phi_{BC}=0$, the red curve does not overlap with the blue curve at its different maxima and minima, unlike the non-thermal scenario. Inside a thermal bath, we found all other features of the transition probability rates to be similar to the non-thermal scenario, and decided not to discuss them.

The broad feature that thermal bath enhances the transition probability rates is also expected to hold with Gaussian switchings. Therefore, for the sake of brevity, we do not investigate the Gaussian switching scenario inside a thermal bath explicitly in the present work.

\section{Discussion}\label{sec:discussion}

In this work, we investigated the radiative process of three quantum probes prepared in a tripartite entangled state. The probes are described by two-level atoms that interact weakly with the background field. We considered the probes to be either static or moving with uniform velocities. We also considered eternal switching and finite Gaussian switching scenarios, as well as the presence of a thermal field bath. In this section, we discuss our key observations, their physical interpretations, and implications. In particular, our main observations are as follows.

\begin{itemize}
    \item In all scenarios corresponding to the eternal switching, our observations suggest that with changing energy gap $(\omega)$ and separation between the probes $(d_{jl})$, the collective transition probability rates do not change monotonically. Rather, they keep oscillating  as these parameters change. For a very low energy gap and separation, the transition probability rate from the symmetric entangled state $|\Omega^{-}_{1}\rangle$ is much larger than the transitions from other states. At the same time, for very large energy gaps and separations, transition probability rates from different entangled states tend to approach similar quantitative values.
    
    \item For static quantum probes and eternal switching, we observed that, corresponding to different probe configurations, the qualitative features of collective transition probability rates from different states can be different. For instance, from Fig. \ref{fig:R-st-et-vOm} we observed that when the probes are configured such that they are on the vertices of an equilateral triangle, the collective transitions from states $|\Omega^{-}_{2}\rangle$, $|\Omega^{-}_{3}\rangle$, and $|\Omega^{-}_{4}\rangle$ are equal. At the same time, when they are placed equidistant on a straight line, the transitions from the states $|\Omega^{-}_{2}\rangle$ and $|\Omega^{-}_{4}\rangle$ are equal. In Fig. \ref{fig:R-st-et-vOm}, we have also plotted the different transition probability rates as functions of the angle $\theta$ between the distance vectors ${\bf d}_{AB}$ and ${\bf d}_{AC}$. These plots indicate that as $\theta$ changes, i.e., as the probe configurations change, so do the features of the transition probability rates from different entangled states. In this regard, one can notice that for $d_{AB}=d_{AC}$, even if $\theta$ changes the transition from $|\Omega^{-}_{2}\rangle$ and $|\Omega^{-}_{3}\rangle$ always remain the same and they become equal with $|\Omega^{-}_{2}\rangle$ only when $\theta=\pi/3$ or $\theta=5\pi/3$, i.e., when the probe configuration corresponds to an equilateral triangle. At the same time, when $d_{AB}=2 d_{AC}$ and $\theta$ changes the transition from all the states in general are different, but the transitions from $|\Omega^{-}_{2}\rangle$ and $|\Omega^{-}_{4}\rangle$ become the same for $\theta=0$, i.e., for the configuration of probes that are equidistant on a straight line. We also observed that in the latter scenario, the transitions from $|\Omega^{-}_{3}\rangle$ and $|\Omega^{-}_{4}\rangle$ states can become equal for certain values of $\theta$. All of these observations indicate that \textbf{the radiative process of tripartite entangled probes can distinguish between certain disparate probe configurations.}

    \item For static probes with Gaussian switching, we observed that the transition probability rates have lesser oscillations as compared to the eternal switching scenario, as $|\omega|d_{jl}$ varies, see Fig. \ref{fig:R-st-Gsn-vOm}. Overall, for finite switching time $T$, the transition probabilities from all the states tend to reach similar quantitative values for lesser separation as compared to the eternal switching. Moreover, we observed that for very low switching times, the transition probability rates from all the entangled states increase with further decrease in $T$. At the same time, for higher $T$, this feature is not universal among different states. Therefore, it suggests that by adjusting the switching time, one can get an enhancement in the transition probability rates corresponding to certain states as compared to others. Whether these features have any practical use in an experimental scenario will be interesting to look forward to.

    \item For probes with inertial motion and eternal switching, our observations suggest that, for the considered parameter values, the transitions from the states $|\Omega^{-}_{1}\rangle$ and $|\Omega^{-}_{4}\rangle$ are the same. At the same time, the transitions from the states $|\Omega^{-}_{2}\rangle$ and $|\Omega^{-}_{3}\rangle$ are also the same. We also observed that when the transition rate from one set of states is at its maxima or minima, the transition from the other set is at minima or maxima, respectively, as the energy gap and the separation between the probes vary. We also observed that depending on the direction of velocity $\phi_{BC}$ of the probes, there can be subtle features in these transition probability rates, distinguishing different scenarios. For instance, when $\phi_{BC}=\pi/2$, i.e., when the probes $B$ and $C$ have velocities perpendicular to their separation, the maxima and minima in transition probability rates corresponding to the different sets of states do not overlap each other, see Fig. \ref{fig:R-vel-et-vOm}. In contrast, when $\phi_{BC}=0$, the maxima and minima in transition probability rates corresponding to different states can overlap. Moreover, from Fig. \ref{fig:R-vel-et-vsv-vsPhi}, we observed that indeed as $\phi_{BC}$ varies, the behaviour of different transition probability rates is opposite to each other. These observations suggest that with tripartite entangled probes, \textbf{one can discern the direction of the probe velocity in addition to different probe arrangements}. It is also noted that as the velocity increases, all the different transition probability rates tend to reach the same quantitative value.

    \item For Gaussian switching and probes in inertial motion, we observed that though the transitions from $|\Omega^{-}_{2}\rangle$ and $|\Omega^{-}_{3}\rangle$ are the same, the transitions from $|\Omega^{-}_{1}\rangle$ and $|\Omega^{-}_{4}\rangle$ are not. The results were obtained here with numerical help, and we observed that as the switching time $T$ grew larger and larger, one can get back results similar to the eternal switching scenario, i.e., the transition rates from $|\Omega^{-}_{1}\rangle$ and $|\Omega^{-}_{4}\rangle$ tend to become equal, see Fig. \ref{fig:R-vel-Gsn-vOm}. Here also, similar to the static probe scenario, the finite switching time lowers the number of oscillations in the transition probability rates as compared to the eternal switching. We also observed that for very low switching time, the transition probability rates from all the entangled states increase with further decrease in $T$. Therefore, one can conclude that even with a non-zero probe velocity, the general features in the transition probability rates due to finite switching remain the same.

    \item \textbf{Thermal bath}: Inside a thermal bath, the transition probability rates are enhanced due to the presence of the thermal bath. This enhancement is in the loss of entanglement or decoherence, which is expected from a noisy environment. The enhancement increases with an increase in temperature, and this also leads to both excitation and de-excitation in the system. When the tripartite entangled system is prepared in the W state $\ket{\Omega^{-}_{n}}$, it can de-excite to the ground state as well as get excited to the second excited state of the collective system state $\ket{\Omega^{+}_{n}}$. The transition rate is higher for de-excitations to the ground state, and as in the case of a quantum field in a vacuum state, the collective transition rates depend on the orientation of the quantum probes. Unlike the vacuum case, in the presence of the thermal state, the transition rate does not follow all the trends of de-excitation and depends on what the initial and final degenerate states are. For example, in an isosceles triangle configuration, the degeneracy for transition rates is different for de-excitation and excitation.
\end{itemize}

We believe our observations are significant in the context of understanding the structure and relative motion between the constituents of a distributed quantum set-up. The transition probability rates for the spontaneous de-excitation of the quantum probes quantify the loss of quantum coherence (decoherence) and entanglement degradation. We have shown that certain entangled states in tripartite entanglement with specific spatial orientation of qubits can inhibit the decoherence effects in a quantum system. Thus, our results can be useful in prescribing a set-up least affected by the effects of quantum decoherence even in a noisy environment. Our set-up will also be helpful in deciphering the contributions due to individual curvature terms of the background spacetime in decoherence - {\textit{quantum gravitational compass}}, a line of thought similar to \cite{K:2023oon}. Moreover, the experimental realization of the W states \cite{Park:2025} makes our work all the more interesting to be implemented in a practical scenario. The prospects of a similar set-up to be prepared in the future space-based gravitational wave detector, Laser Interferometer Space Antenna or LISA, do not seem impossibly far-fetched.

\begin{acknowledgments}

The authors would like to thank Dawood Kothawala for discussions and comments on the manuscript. K. H. would like to thank the Indian Institute of Technology Bombay (IIT Bombay) for supporting this work through a postdoctoral fellowship.

\end{acknowledgments}

\widetext 

\appendix

\section{Evaluation of $F_{jj}(\omega)$ for static probes with Gaussian switching}\label{Appn:Fjj-Inertial-Gaussian}
In this section of the Appendix, we evaluate the expression of $F_{jj}(\omega)$ for static probes interacting with the background field with Gaussian switching. In this regard, we consider the definition of $F_{jj}(\omega)$ from Eq. \eqref{eq:response-fn} and the Wightman function for static probes from Eq. \eqref{eq:SP-Gjj}. Then with Gaussian switching $\kappa(\tau_{j})=e^{-\tau_{j}^2/T^2}$, we have
\begin{eqnarray}\label{eq:SP-Fjj-Gaussian-1}
    F_{jj}(\omega) = -\frac{1}{8\pi^2}\int_{-\infty}^{\infty}ds_{j}\int_{-\infty}^{\infty}du_{j}\,\frac{e^{-i\,\omega u_{j}}}{(u_{j}-i\epsilon)^2}\,e^{-(u_{j}^2+s_{j}^2)/(2T^2)}
\end{eqnarray}
In the previous expression we have utilized a change of variables $s_{j}=\tau_{j}+\tau'_{j}$ and $u_{j}=\tau_{j}-\tau'_{j}$. In the subsequent analysis we will be using certain integral representations which are as follows.
\begin{subequations}\label{eq:Gen-IntRep-1}
\begin{eqnarray}\label{eq:Gen-IntRep-1a}
    \int_{-\infty}^{\infty}ds\,e^{-s^2/(2T^2)} &=& T\,\sqrt{2\pi}~,\\
    \textup{and}~~ e^{-u^2/(2T^2)} &=& \frac{T}{\sqrt{2\pi}}\, \int_{-\infty}^{\infty} d\xi\,e^{i\,\xi\,u-\xi^2\,T^2/2}~.\label{eq:Gen-IntRep-1b}
\end{eqnarray}
\end{subequations}
Then with the help of Eq. \eqref{eq:Gen-IntRep-1a} and the Fourier representation of \eqref{eq:Gen-IntRep-1b}, the auto-correlation function $F_{jj}$ can be expressed as
\begin{eqnarray}\label{eq:SP-Fjj-Gaussian-2}
    F_{jj}(\omega) &=& -\frac{T^2}{8\pi^2}\int_{-\infty}^{\infty}d\xi\,e^{-\xi^2\,T^2/2}\int_{-\infty}^{\infty}du_{j}\,\frac{e^{i\,(\xi-\omega)\,u_{j}}}{(u_{j}-i\epsilon)^2}~\nonumber\\
    ~&=& \frac{T^2}{4\pi}\int_{\omega}^{\infty}d\xi\,(\xi-\omega)\,e^{-\xi^2\,T^2/2}~\nonumber\\
    ~&=& \frac{1}{4\pi}\,\Big[e^{-\omega^2\,T^2/2}-\sqrt{\frac{\pi}{2}}\,T\,\omega\, \erfc\left( \frac{\omega \,T}{\sqrt{2}}\right)\Big]~.
\end{eqnarray}
In the above the first integral over $u_{j}$ is carried out using the residue theorem, where the integrand has a pole of order two at $u_{j}=i\epsilon$. The second integral over $\xi$ is also straightforward to evaluate. One should note that the quantity $F_{jj}(\omega)$ corresponds to the individual probe transition probability, and this result corresponding to the Gaussian switching is known for quiet some time, see \cite{LSriramkumar_1996, Xu:2020pbj, Barman:2023aqk, K:2023oon, Barman:2024vah}.

\section{Evaluation of $F_{BB}(\omega)$ for probes in uniform velocity and with eternal switching}\label{Appn:FBB-Inertial-Eternal}

Here, we consider the situation where the atomic probe $B$ is in uniform velocity, and it interacts with the background field through an eternal switching, i.e., $\kappa(\tau)=1$. The transition coefficient $F_{BB}(\omega)$ can obtained as
\begin{eqnarray}\label{AppnEq:UVP-FBB-eternal-1}
    F_{BB}(\omega) &=& -\frac{1}{4\pi^2} \int_{-\infty}^{\infty} d\tau'_{B} \int_{-\infty}^{\infty} d\tau_{B} \frac{e^{-i\,\omega(\tau_{B}-\tau'_{B})}}{\left\{\gamma_{B}(\tau_{B}-\tau'_{B})-i\,\epsilon\right\}^2-\gamma_{B}^2 \textsc{v}_{B}^2(\tau_{B}-\tau'_{B})^2}\nonumber\\
    &=& -\frac{1}{8\pi^2} \int_{-\infty}^{\infty} ds_{B} \int_{-\infty}^{\infty} du_{B} \frac{e^{-i\,\omega\,u_{B}}}{\{\gamma_{B}(1+\textsc{v}_{B})u_{B}-i\,\epsilon\}\,\{\gamma_{B}(1-\textsc{v}_{B})u_{B}-i\,\epsilon\}}~.
\end{eqnarray}
In the above expression, we considered change of variables  $s_{B}=\tau_{B}+\tau'_{B}$ and $u_{B}=\tau_{B}-\tau'_{B}$, for ease of calculation. One can notice that the integrand, after this change of variables, depends only on the variable $u_{B}$. The above integration becomes
\begin{eqnarray}\label{AppnEq:UVP-FBB-eternal-2}
    F_{BB}(\omega) &=& \frac{1}{8\pi^2}\,\frac{1}{2i\,\epsilon\,\textsc{v}_{B}} \int_{-\infty}^{\infty} ds_{B} \int_{-\infty}^{\infty} du_{B} \bigg[\frac{(1+\textsc{v}_{B})\,e^{-i\,\omega\,u_{B}}}{\gamma_{B}(1+\textsc{v}_{B})u_{B}-i\,\epsilon}- \frac{(1-\textsc{v}_{B})\,e^{-i\,\omega\,u_{B}}}{\gamma_{B}(1-\textsc{v}_{B})u_{B}-i\,\epsilon}\bigg]\nonumber\\
    &=& \lim_{\epsilon\to 0}~ \frac{1}{8\pi^2}\,\frac{1}{2i\,\epsilon\,\textsc{v}_{B}} \int_{-\infty}^{\infty} ds_{B} \frac{\Theta(-\omega)2\pi\,i}{\gamma_{B}}\,\Big[e^{\epsilon\,\omega/\{\gamma_{B}(1+\textsc{v}_{B})\}}-e^{\epsilon\,\omega/\{\gamma_{B}(1-\textsc{v}_{B})\}}\Big] \nonumber\\
    &=& -\frac{\omega\,\Theta(-\omega)}{4\pi}\, \int_{-\infty}^{\infty} ds_{B}~.
\end{eqnarray}
We would like to mention that one can obtain the quantity $F_{CC}(\omega)$ in a similar fashion.

\section{Evaluation of $F_{BB}(\omega)$ for probes in uniform velocity and with Gaussian switching}\label{Appn:FBB-Inertial-Gaussian}

In this part of the Appendix, we consider that the probe $B$ is in uniform velocity, and it interacts with the background field through a finite Gaussian switching, i.e., the window function is given by $\kappa(\tau)=e^{-\tau^2/T^2}$. We take the expression of $F_{jj}(\omega)$ from Eq. \eqref{eq:response-fn}, and for probe $B$ this expression results in
\begin{eqnarray}\label{AppnEq:UVP-FBB-gaussian-1}
    F_{BB}(\omega) &=& -\frac{1}{8\pi^2}\,\int_{-\infty}^{\infty} ds_{B}\int_{-\infty}^{\infty} du_{B}\,\frac{e^{-i\,\omega\,u_{B}}\,e^{-(u^2_{B}+s^2_{B})/(2T^2)}}{\{\gamma_{B}(1+\textsc{v}_{B})u_{B}-i\,\epsilon\}~\{\gamma_{B}(1-\textsc{v}_{B})u_{B}-i\,\epsilon\}}~\nonumber\\
    ~&=& \frac{T^2}{8\pi^2}\,\frac{1}{2\,i\,\epsilon\,\textsc{v}_{B}\,\gamma_{B}}\, \int_{-\infty}^{\infty} d\xi\, e^{-\xi^2 T^2/2}\,\int_{-\infty}^{\infty}du_{B}\, e^{i\,(\xi-\omega)u_{B}}\,\bigg[\frac{1}{u_{B}-\frac{i\,\epsilon}{\gamma_{B}(1+\textsc{v}_{B})}}-\frac{1}{u_{B}-\frac{i\,\epsilon}{\gamma_{B}(1-\textsc{v}_{B})}}\bigg]~.
\end{eqnarray}
In Eq. \eqref{AppnEq:UVP-FBB-gaussian-1}, we have considered a change of variables $s_{B}=\tau_{B}+\tau'_{B}$ and $u_{B}=\tau_{B}-\tau'_{B}$. Moreover, to obtain the last expression we have utilized the integral representations from Eq. \eqref{eq:Gen-IntRep-1}.
We employ the residue theorem to evaluate the integral of Eq. \eqref{AppnEq:UVP-FBB-gaussian-1}. In particular, after a few simple algebras we obtain
\begin{eqnarray}
    F_{BB}(\omega) &=& \lim_{\epsilon\to 0} \frac{T^2}{8\pi}\,\frac{1}{\epsilon\,\textsc{v}_{B}\,\gamma_{B}}\, \int_{-\infty}^{\infty} d\xi\, e^{-\xi^2 T^2/2}\,\Theta(\xi-\omega)\,\bigg[\exp{\left\{-\frac{(\xi-\omega)\,\epsilon}{\gamma_{B}(1+\textsc{v}_{B})}\right\}}-\exp{\left\{-\frac{(\xi-\omega)\,\epsilon}{\gamma_{B}(1-\textsc{v}_{B})}\right\}}\bigg]~\nonumber\\
    ~&=& \frac{T^2}{4\pi}\, \int_{\omega}^{\infty} d\xi\, e^{-\xi^2 T^2/2}\,(\xi-\omega)~\nonumber\\
    ~&=& \frac{1}{4\pi}\, \bigg[e^{-\omega^2 T^2/2}-T\,\omega\sqrt{\frac{\pi}{2}}\,\erfc\left( \frac{T\,\omega}{\sqrt{2}}\right)\bigg]~.
\end{eqnarray}
Here also, we would like to mention that the quantity $F_{CC}$ can be obtained in a similar manner.

\section{Evaluation of $F_{BC}$ for probes in uniform velocity and with Gaussian switching}\label{Appn:FBC-Inertial-Gaussian}

To obtain the final expression of $F_{BC}$, where both the probes $B$ and $C$ are in uniform velocity, we consider its initial expression from Eq. \eqref{eq:UVP-Fbc-Gaussian-1}. We also note that we have considered a specific scenario where $\textbf{\textsc{v}}_{B}= \textbf{\textsc{v}}_{C}= \textbf{\textsc{v}}$, and thus $\gamma_{B}=\gamma_{C}=\gamma$. By integration over $u_{BC}$, $\xi$, and $s_{BC}$ chronologically we obtain the final result as
\begin{eqnarray}\label{eq:UVP-Fcb-Gaussian-2}
    F_{BC}(\omega) &=& \frac{T}{8 d_{BC}\sqrt{2\pi } \sqrt{1-\textsc{v}^2 \sin^2(\phi_{BC})}} \nonumber\\
    ~&& \exp \bigg[\tfrac{2 d_{BC}  \sqrt{1-\textsc{v}^2 \sin^2(\phi_{BC})} \big\{d_{BC} \textsc{v} \cos (\phi_{BC})+i T^2 \omega/\gamma\big\}+d_{BC} \big\{d_{BC} \textsc{v}^2 \cos (2 \phi_{BC})+d_{BC}+2 i T^2 (\textsc{v}/\gamma) \omega \cos (\phi_{BC})\big\}}{-(2 T^2/\gamma^2)}\bigg] \nonumber\\
    ~&& \Bigg[i \left\{\frac{1}{\gamma}- \,\frac{1}{\gamma} \erf\left(\frac{- d_{BC}\gamma  \sqrt{1-\textsc{v}^2 \sin^2(\phi_{BC})}+ d_{BC}\gamma \textsc{v} \cos (\phi_{BC})+i T^2 \omega}{\sqrt{2}\,i\, T }\right)\right\} \nonumber\\
    ~&&\exp \left\{\frac{i 2 d_{BC}\gamma \sqrt{1-\textsc{v}^2 \sin^2(\phi_{BC})} \left(T^2 \omega-i d_{BC} \gamma \textsc{v} \cos (\phi_{BC})\right)}{T^2}\right\}\nonumber\\
    ~&& +\,\frac{i}{\gamma} \,\erf\left(\gamma\frac{\sqrt{2} d_{BC}  \sqrt{1-\textsc{v}^2 \sin^2(\phi_{BC})}+\sqrt{2} d_{BC} \textsc{v} \cos (\phi_{BC})+i T^2 \sqrt{2-2 \textsc{v}^2} \omega}{2\,i\, T }\right)-i \frac{1}{\gamma}\Bigg]\,= \bigg(T\sqrt{\frac{\pi}{2}}\bigg)R_{BC}(\omega)~.\nonumber\\
\end{eqnarray}
In this expression of $R_{BC}(\omega)$ if one takes the velocity of the probes to be vanishing, i.e., if one considers $\textsc{v}\to 0$, one should get the expression of $R_{BC}(\omega)$ corresponding to the static probe scenario from Eq. \eqref{eq:SP-Fjl-Gaussian-3}. Let us verify this limit. In particular, we will be using the \emph{Error} function identities $\erfc(x)=1-\erf(x)$, $\mathrm{Erfi}(x)=-i\,\erf(i\,x)$, $\erf(-x)=-\erf(x)$. Then considering the expression of Eq. \eqref{eq:UVP-Fcb-Gaussian-2} we get
\begin{eqnarray}\label{eq:UVP-Fcb-Gaussian-3}
    \lim_{v\to 0} R_{BC}(\omega) &=& \frac{e^{-\frac{1}{2} d_{BC} \left(\frac{d_{BC}}{T^2}+2 i \omega \right)}}{8 \pi  d_{BC}}\, \left\{i e^{2 i d_{BC} \omega } \erfc\left(\frac{T^2 \omega +i d_{BC}}{\sqrt{2} T}\right)+\mathrm{Erfi}\left(\frac{d_{BC}+i T^2 \omega }{\sqrt{2} T}\right)-i\right\}\nonumber\\
    ~&=& -\frac{i\, e^{-\frac{1}{2} d_{BC} \left(\frac{d_{BC}}{T^2}+2 i \omega \right)}}{8 \pi  d_{BC}}\, \left\{\erfc\left(\frac{T^2 \omega -i d_{BC}}{\sqrt{2} T}\right)-e^{2 i \omega  d_{BC}} \erfc\left(\frac{T^2 \omega +i d_{BC}}{\sqrt{2} T}\right)\right\}~,
\end{eqnarray}
which is the same as the expression of $R_{BC}(\omega)$ as obtained in Eq. \eqref{eq:SP-Fjl-Gaussian-3} corresponding to the static probe scenario.

\section{$R_{BC}$ for probes in uniform velocity with eternal switching inside a thermal bath}\label{Appn:RBC-Th-Inertial}
We consider the Wightman function of Eq. \eqref{eq:TH-inertial-intr-trans} corresponding to three probes inside a thermal bath. In this set-up the probe $A$ is static. At the same time, probe $B$ and $C$ are in uniform velocity and $\textbf{\textsc{v}}_{B}=\textbf{\textsc{v}}_{C}=\textbf{\textsc{v}}$. With this probe configuration, we obtain the transition probability rate with the help of Eq. \eqref{eq:TH-UVP-Fbc-eternal-2} for $\omega<0$ as
\begin{eqnarray}\label{eq:RBC-Th-Vel-Oml0}
    R_{BC}(\omega) &=& -\sum_{n=0}^{\infty}\frac{\exp{\big[\omega \big\{\beta  n-i d_{BC} \textsc{v}\, \cos (\phi_{BC} )\big\}/\{\gamma (1-\textsc{v}^2)\}}\big]}{8 \pi  \sqrt{\gamma^2 \big[-d_{BC}^2 \left(\textsc{v}^2-1\right)+d_{BC} \textsc{v}\, \cos (\phi_{BC} ) \big\{d_{BC} \textsc{v}\, \cos (\phi_{BC} )+2 i \beta  n\big\}-\beta ^2 n^2 \textsc{v}^2\big]}}\nonumber\\
    ~&\times& \sin \left(\omega \sqrt{\gamma^2 \big[-d_{BC}^2 \left(\textsc{v}^2-2\right)+d_{BC} \textsc{v}\, \big\{d_{BC} \textsc{v}\, \cos (2 \phi_{BC} )+4 i \beta  n \cos (\phi_{BC} )\big\}-2 \beta ^2 n^2 \textsc{v}^2\big]}/\big\{\sqrt{2} \gamma^2 (1-\textsc{v}^2)\big\}\right)~.\nonumber\\
\end{eqnarray}
In the zero temperature scenario, one can take the limit $\beta\to 0$, which is the same as taking only the $n=0$ term of the above sum, to obtain the transition probability rate $R_{BC}(\omega)$. The zero temperature $R_{BC}(\omega)$ obtained from Eq. \eqref{eq:RBC-Th-Vel-Oml0} in this manner is the same as the one from Eq. \eqref{eq:UVP-Rcb-eternal}. At the same time, for $\omega>0$, the entire transition probability rate is obtained from the sum
\begin{eqnarray}\label{eq:RBC-Th-Vel-Omg0}
    R_{BC}(\omega) &=& \sum_{n=-\infty}^{-1}\frac{\exp{\big[-\omega \big\{\beta  n-i d_{BC} \textsc{v}\, \cos (\phi_{BC} )\big\}/\{\gamma (1-\textsc{v}^2)\}}\big]}{8 \pi  \sqrt{\gamma^2 \big[-d_{BC}^2 \left(\textsc{v}^2-1\right)+d_{BC} \textsc{v}\, \cos (\phi_{BC} ) \big\{d_{BC} \textsc{v}\, \cos (\phi_{BC} )+2 i \beta  n\big\}-\beta ^2 n^2 \textsc{v}^2\big]}}\nonumber\\
    ~&\times& \sin \left(-\omega \sqrt{\gamma^2 \big[-d_{BC}^2 \left(\textsc{v}^2-2\right)+d_{BC} \textsc{v}\, \big\{d_{BC} \textsc{v}\, \cos (2 \phi_{BC} )+4 i \beta  n \cos (\phi_{BC} )\big\}-2 \beta ^2 n^2 \textsc{v}^2\big]}/\big\{\sqrt{2} \gamma^2 (1-\textsc{v}^2)\big\}\right)~.\nonumber\\
\end{eqnarray}
In the sum of expression \eqref{eq:RBC-Th-Vel-Omg0}, as all $n$ are negative, the transition probability rate will vanish as one takes the limit $\beta\to \infty$. This observation signifies that without the presence of a thermal bath, there will be no excitations in the atomic system, which is consistent with the results from our analysis of the non-thermal scenario. We would also like to mention that the sums in Eqs. \eqref{eq:RBC-Th-Vel-Oml0} and \eqref{eq:RBC-Th-Vel-Omg0} cannot be evaluated, up to our understanding, analytically and should be estimated numerically.

\end{document}